\documentclass[apj]{emulateapj}
\usepackage{multirow}
\usepackage{graphics}
\usepackage{rotating}
\usepackage{enumitem}
\usepackage{hyperref}
\hypersetup{
  hyperfootnotes=true,    % footnote marks are links to footnote text
  colorlinks=true,        % false: boxed links; true: colored links
  linkcolor=blue,         % color of internal links
  citecolor=blue,         % color of links to bibliography
  urlcolor=blue,          % color of external links
  breaklinks=false,        % allow links to break over lines
}

\shorttitle{Demographics of Isolated Galaxies}
\shortauthors{Khim et al.}

\def\kms{${\rm km}~{\rm s}^{-1}$}

\newcommand{\gandalf}{{\texttt {gandalf}}}
\newcommand{\fracDeV}{{\texttt {FracDeV}}} 
\newcommand{\zobov}{{\texttt {ZOBOV}}} 
\newcommand{\kcorrect}{{\texttt {KCORRECT}}}

\def\OIII{[\mbox{O\,{\sc iii}}]~$\lambda 5007$ }
\def\NII{[\mbox{N\,{\sc ii}}]~$\lambda 6584$ }

\begin{document}

\title{Demographics of Isolated Galaxies Along the Hubble Sequence}

\author{Hong-geun Khim\altaffilmark{1}, Jongwon Park\altaffilmark{1}, Seong-Woo Seo\altaffilmark{1}, Jaehyun Lee\altaffilmark{1}, Rory Smith\altaffilmark{1} and Sukyoung K. Yi\altaffilmark{1,2}}
\altaffiltext{1}{Department of Astronomy, Yonsei University, Seoul 120-749, Korea; yi@yonsei.ac.kr} 
\altaffiltext{2}{Yonsei University Observatory, Yonsei University, Seoul 120-749, Korea}

\begin{abstract}

Isolated galaxies in low-density regions are significant in the sense that they are least affected by the hierarchical pattern of galaxy growth, and interactions with perturbers, at least for the last few Gyr. To form a comprehensive picture of the star formation history of isolated galaxies, we constructed a catalog of isolated galaxies and their comparison sample in relatively denser environments. The galaxies are drawn from the SDSS DR7 in the redshift range of $0.025<z<0.044$. We performed a visual inspection and classified their morphology following the Hubble classification scheme. For the spectroscopic study, we make use of the OSSY catalog. We confirm most of the earlier understanding on isolated galaxies. The most remarkable additional results are as follows. Isolated galaxies are dominantly late type with the morphology distribution (E: S0: S: Irr) = (9.9: 11.3: 77.6: 1.2)\%. The frequency of elliptical galaxies among isolated galaxies is only a third of that of the comparison sample. Most of the photometric and spectroscopic properties are surprisingly similar between isolated and comparison samples. However, early-type isolated galaxies are less massive by 50\% and younger (by H$\beta$) by 20\% than their counterparts in the comparison sample. This can be explained as a result of different merger and star formation histories for differing environments in the hierarchical merger paradigm. We provide an on-line catalog for the list and properties of our sample galaxies.
\end{abstract}


% --------------------------------------------------------------------------------

\section{Introduction}
The differing properties of galaxies in high- and low-density regions indicate that there is a tug-of-war between the influence of internal processes and external factors on galaxy evolution, depending on the surroundings \citep{bla05,van08,pen10,fer13}. In the higher density environment, galaxies are subjected to external effects such as interactions or mergers which seem frequent and universal in the densely populated system, leading to enhanced star formation activity compared to field galaxies \citep{san96,bar00,lam03,alo04,nik04,jog09}. In addition, major mergers can deform the morphology of a galaxy \citep{too72,sch92,naa99} causing a higher fraction of early-type galaxies in high-density environments \citep{oem74,dre80,tre03}.

On the other hand, the galaxies in less dense regions, like field galaxies, are thought to have experienced a relatively small number of mergers. This low merger rate helps internal processes become significant because internal secular evolution, such as the formation of galactic bars and spiral structures, is too slow and fragile to be sustained against violent major mergers \citep{kor04}. Some studies, in this less harsh environment, found that early-type galaxies are observed to be less massive and relatively younger \citep{bau96,kau98,gov99,col00,kun02}, and late-type galaxies prefer to have some internally developed features such as a pseudo-bulge \citep{kor04,dro07,fis08}, confirming the differing evolution of galaxies in low-density regions.

An isolated galaxy is considered as an object which has not been considerably perturbed by its neighbors for a few Gyr \citep{sto78,all05,ver05}. Therefore, in the context of the local environment effect, many studies of isolated galaxies have been conducted in the last few decades \citep{tur75,bal81,vet86,zar93,mar96,mar99,aar01,col01,mar02,pis02,mar03,pra03,var04}. One of the well known studies is the Analysis of the interstellar Medium of Isolated GAlaxies (AMIGA) project whose galaxies are based on the Catalogue of Isolated Galaxies \citep[CIG;][]{kar73}. The AMIGA project refined the CIG for morphologies of 1,050 isolated galaxies \citep{sul06} using the second Palomar Observatory Sky Survey (POSS II). The degree of isolation was also quantified \citep{ver07b,ver07c} using the digitized POSS (DPOSS I and II). In a series of studies, the AMIGA project dealt with multiwavelength properties of isolated galaxies including photometric characteristics \citep{ver05,sul06,dur08,dur09,fer12,fer13}, the star formation rate derived from H$\alpha$ emission \citep{ver07a}, the FIR luminosity function \citep{lis07}, radio continuum properties \citep{leo08} and the nuclear activity using FIR, radio and optical databases \citep{sab08,sab12}.

In defining an isolated galaxy, the isolation criteria devised by \citet{kar73}, or variations on this theme, are widely used in studies including AMIGA projects. \citet{kar73} only took into account a relative apparent size and a projected distance between the target galaxy and potential neighbors, due to the deficit of redshift information at that time. But the photometric isolation scheme can miss some isolated galaxies by regarding background objects as potential companions \citep{ver07c,arg13}.

In addition to this, tidal perturbation strength can be considered as a criterion of isolation \citep{ver07b,ver07c}. The tidal perturbation strength helps to quantify the tidal influence of the companions based on spatial separations and stellar mass ratio between the target galaxy and companions. For example, if external tidal forces are greater than 1\% of the internal binding force, the galaxy can be thought to be perturbed by neighbors \citep{ath84,var04}. However, the external influence may not be instantaneous but continuous, so accumulated effects during the transit time of neighboring galaxies should be considered to better understand the effects of perturbers.

Statistical studies of isolated galaxies using the Sloan Digital Sky Survey \citep[SDSS;][]{yor00} have been reported \citep{all05,her10,arg13}. Many of them are based on the criteria developed by \citet{kar73}, and only focused on providing a catalog of isolated galaxies or testing the CIG criteria.

In this paper, we constructed a catalog of isolated galaxies using the SDSS Data Release 7 \citep[DR7;][]{aba09} to present the properties of isolated galaxies along the Hubble morphological type. For the isolation criteria, we took into account the distance in three dimensional space to reduce the impact of projection effects. We did not use the tidal perturbation strength, but instead rely on a strict distance criterion. In Section~\ref{sec:sample}, we present our sample selection criteria for isolated galaxies and comparison sample galaxies. In Section~\ref{sec:basic}, we describe the morphologcal classification scheme and photometric properties of isolated galaxies. The characteristics of absorption lines with a stellar population model and the properties of emission lines are addressed in Sections~\ref{sec:absorption} and~\ref{sec:emission}, respectively. In section~\ref{sec:discussion}, we discuss the evolutionary scenario of isolated galaxies compared to comparison galaxies and possible biases in the sampling. We present our conclusions and summarize our results in Section~\ref{sec:summary}. Throughout this work, we have assumed the following cosmological parameters: H$_0$ = 70 \kms Mpc$^{-1}$, $\Omega_{\rm M}$ = 0.3 and $\Omega_\Lambda$ = 0.7.

% Figure - Spatial distribution
\begin{figure*}
\centering
\includegraphics[width=0.48\textwidth, angle=270]{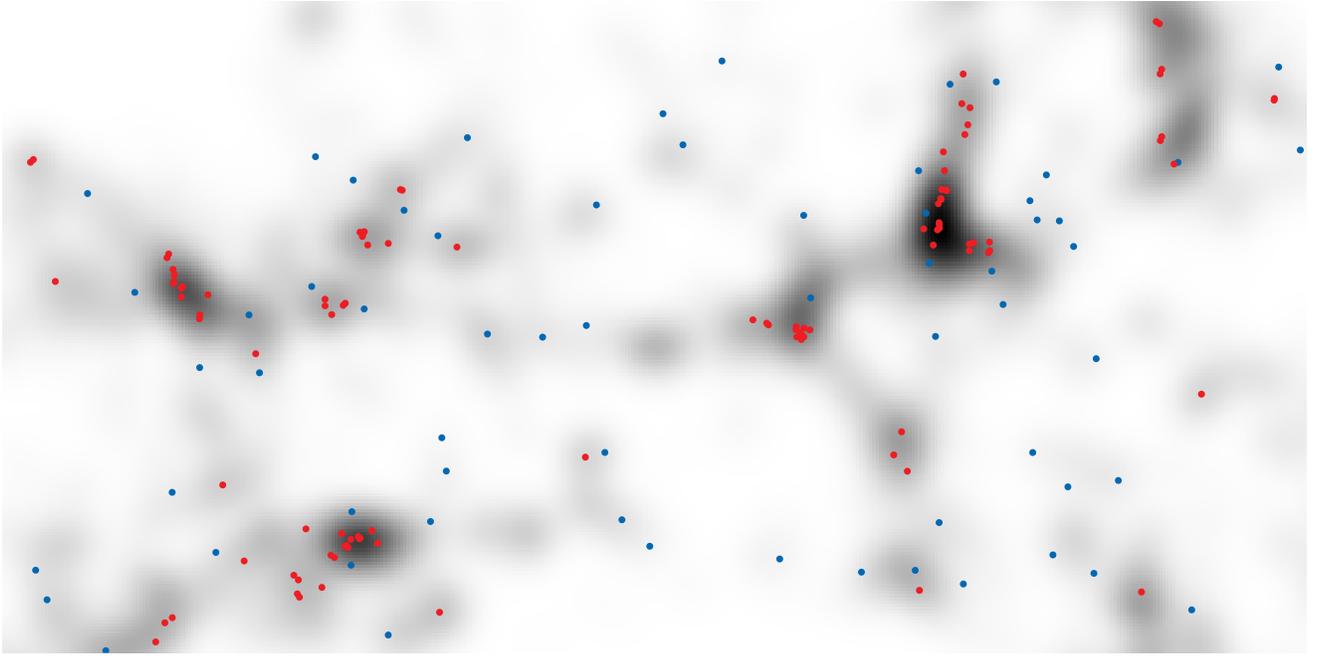}
\caption{A projected (column) density map with 200 x 100 x 5 Mpc. The 5 Mpc depth is chosen to minimize the projection effect. Gray scale shows long-range galaxy densities derived using the Gaussian weighted method and darker colors indicate higher densities. Blue and red dots correspond to isolated and comparison sample galaxies, respectively.}
\label{fig:spatial}
\end{figure*}

% --------------------------------------------------------------------------------

\section{Sample Selection}
\label{sec:sample}
To study isolated galaxies, we make two kinds of samples: isolated galaxies without any companions, and a comparison sample of galaxies in denser regions. The parent sample consists of the main galaxy sample \citep{str02} of the SDSS DR7 which covers one quarter of the sky using a 2.5m telescope in Apache Point, New Mexico \citep{gun06}. We focus on the nearby galaxies ($0.025 < z < 0.044$) so as to include more faint galaxies when we measure the galaxy density in our volume-limited sample. This redshift range also helps us to secure relatively robust visual morphological classification while minimizing the evolution effect, as described by \citet{oh13}.

The main galaxy sample is comprised of galaxies with spectroscopic information and most of them are at $z \lesssim 0.2$. The sample is designed to be complete for galaxies with $r \leq 17.77$ in spectroscopy. However, galaxies with low surface brightness or galaxies which are too bright in apparent fiber magnitudes are not targeted for the spectroscopic survey \citep{str02}. Furthermore, some galaxies in crowded fields can be missed due to the fiber collisions \citep{van07,yoo08}. Therefore, before defining isolated galaxies, we test the spectroscopic completeness for each field of our galaxies. The spectroscopic completeness of each field is estimated by calculating the percentage of galaxies with spectra among galaxies in the luminosity range of $13.0 \leq r \leq 17.77$, a projected 2 Mpc radius around our target galaxy. 2 Mpc is our criterion of isolation (see below). Since the spectroscopic completeness declines for bright galaxies, we set the minimum $r$-band magnitude to be 13.0 which guarantees at least 70\% of the spectroscopic completeness \citep{mon09}. We have found that only about 10\% of galaxies in $0.025 < z < 0.044$ (with the mean completeness of 89\%) have the completeness less than 70\% and they are removed at this stage.

The isolated galaxies and the comparison group galaxies used in our study are selected based on a volume-limited sample ($M_{\rm r} \leq -18.68$). This magnitude cut roughly corresponds to a stellar mass of $10^{9}\ M_\odot$. After the volume limitation, we have 38,011 galaxies.
 
For this volume-limited sample, we measure the density of each galaxy by counting the number of companion galaxies within a sphere in three dimensional space. In estimating a distance between galaxies using the redshift, the distance could be overestimated due to proper motions of field galaxies \citep[$\sim$190 \kms;][]{ton00}, misclassifying some galaxies to be isolated. In other studies, an elongated cylinder with 0.5 or 1 Mpc radius is considered to compensate for the effect of the proper motion in selecting isolated galaxies \citep{ver07c,arg15}. Since we define isolated galaxies based on a sphere instead of a cylinder, a 2 Mpc distance criterion in redshift space, which is a strict condition, is chosen. This 2 Mpc criterion can help companion galaxies to be separated from the target galaxy by at least 1 Mpc in ``real'' space. For example, in the worst case scenario, the target and companion galaxy are aligned along the line of sight. We take the case that the galaxies appear separated by 2 Mpc in redshift space, but in fact the companion galaxy has a proper motion of 190 \kms. We calculate that if we vary the direction of the proper motion of the companion randomly, the companion will have a true separation from the target galaxy greater than 1 Mpc, 70\% of the time (prob = $\frac{\Delta \Omega}{4\pi}$ = $\frac{1}{2}\int_{7\pi/18}^{\pi} sin(\theta)d\theta$ = 0.7). As this is a worst case scenario, we can expect our 2 Mpc criterion to guarantee at least a 1 Mpc true separation in most cases.

To choose the comparison galaxies as the ``typical'' galaxies according to the hierarchical formation scenario, we select the comparison galaxies when there are at least two companion galaxies in a 1 Mpc sphere under the consideration of the finger-of-God effect  \citep{sch07b,yoo08}. A typical system in the comparison sample is similar to the Local Group -- the Milky way has two companion galaxies within 1 Mpc, M31 and M33.

When we account for companion galaxies, the mass ratio between the target galaxy and companion galaxies varies depending on the mass of the target galaxy. For example, massive galaxies can have a large range of mass ratio. Meanwhile, for galaxies near the magnitude limit ($M_{\rm r}\sim -18.68$), only galaxies with comparable mass can be considered as companion galaxies. In other words, our definition of isolation could permit some ``isolated galaxies'' to have companion galaxies with comparable mass below the magnitude cut off. To minimize this inhomogeneity in the sampling, we select galaxies with $M_{\rm r} \leq -19.68$ (1 magnitude brighter than the magnitude limit). This ensures companions can be detected near the magnitude limit down to mass ratio of $\sim$1:3 (the mass limit of target galaxies is $\sim$$10^{9.5}$ $M_\odot$ and that of companion galaxies is $\sim$$10^{9}$ $M_\odot$). With this choice, the number of isolated galaxies and comparison sample galaxies are 3,969 and 4,591, respectively.

% --------------------------------------------------------------------------------

\section{Basic properties of galaxies}
\label{sec:basic}
\subsection{Spatial Distribution}
To examine the surroundings of our galaxies, it is helpful to visualize the spatial distribution of the galaxies. Figure~\ref{fig:spatial} shows a column density map with isolated and comparison sample galaxies. We measure the Gaussian weighted density of galaxies for our volume-limited sample (38,011) \citep{sch07b,yoo08}:

\begin{equation}
\rho(\sigma) = \frac{1}{\sqrt{2\pi}\sigma}exp\left(-\frac{1}{2}\left(\frac{r^2_a}{\sigma^2}+\frac{r^2_z}{c_z^2\sigma^2}\right)\right)
\end{equation}

\noindent $r_a$ is the angular distance and $r_z$ is the distance along the line of sight. $c_z$ is introduced to compensate for the finger-of-god effect. We choose $\sigma$ to be 3 Mpc to present long-range densities. The darker shading indicates the higher long-range density and isolated and comparison sample galaxies are shown in blue and red symbols, respectively.

A significant fracton of the volume is empty space (voids). Relatively denser regions form filamentary structures. Most of comparison sample galaxies seem to reside in the vicinity of the high-density region. Meanwhile, isolated galaxies are widely distributed from relatively less dense regions to higher density areas \citep{arg14,arg15}. It is clear that the definition of isolated galaxies in our sample is with respect to the local density, rather than the global environment.

It is interesting to compare our galaxies with void galaxies which are regarded as galaxies in very low-density regions. For a void catalog, we adopted the catalog provided by \cite{sut12}. \cite{sut12} tried to identify voids in SDSS DR 7 using a void finder $\zobov$ which is based on the Voronoi tessellation. The $\zobov$ algorithm is liable to include the surrounding high-density walls of the void \citep{ney08}. Therefore, to avoid including the galaxies in void walls, we only select the members located in the central regions of voids by using smaller effective size of voids than the given value, i.e. 0.8R$_{\rm eff}$. As a result, we find 59 voids in our sampling volume with a median radius of 7 Mpc, and  $\sim$3\% of isolated galaxies and $\sim$4\% of comparison galaxies are in voids. Comparable numbers of void galaxies in the isolated and comparison sample indicates that the void galaxies can be either isolated or coupled. Also, a small number of void galaxies in our sample implies that void galaxies are in ``globally'' low-density regions while our isolated galaxies are in ``locally'' low-density areas. Thus the isolated galaxies can be expected to show properties caused by local environments, rather than global ones.

% --------------------------------------------------------------------------------

\subsection{Morphology Classification}
\label{sec:morph}

It is hard to classify the morphology of galaxies that appear small or edge-on. Thus, we take advantage of the criteria used in \citet{oh13} to remove apparently small or low IsoB$_{\rm r}$/IsoA$_{\rm r}$ objects among isolated and comparison sample galaxies: IsoA$_{\rm r}$ $<$ 30$\arcsec$ or IsoB$_{\rm r}$/IsoA$_{\rm r}$ $<$ $-$0.01 IsoA$_{\rm r}$+0.90. Here, IsoA$_{\rm r}$ and IsoB$_{\rm r}$ from the SDSS photometric pipeline are the isophotal radius in $r$-band along the major-axis and minor-axis, respectively. To test the criteria, we also perform a morphological classification using a randomly selected 500 galaxies from our sample, and confirm that these demarcation lines separate the classifiable and non-classifiable objects well. This step removes about 30\% of galaxies, leaving 2,660 isolated galaxies and 3,409 comparison sample galaxies for a visual inspection. When we exclude the galaxies which cannot satisfy the criteria, some galaxies with intrinsically low IsoB$_{\rm r}$/IsoA$_{\rm r}$ like E6 and E7 are ruled out, making a bias against them. Therefore, we will not address such galaxies in our study.

3 $gri$ composite color images with different scales, and one gray scale image from the SDSS are used in the visual inspection. The field-of-view of the 3 colored images are 50, 70 and 100 kpc. We classify the morphology as elliptical, lenticular, unbarred spiral, barred spiral and irregular galaxy following the Hubble sequence \citep{hub26}. Galaxies which cannot be placed in a specific class are categorized as ``unknown'' type galaxies. We did not consider other revised versions of Hubble types such as classifications used in Third Reference Catalog of Bright Galaxies \citep[RC3;][]{deV91} to keep simplicity.

The morphology classification strategy we use is similar to the classification method described in \citet{oh13}. Here is a brief description of the scheme.

\begin{enumerate}[leftmargin=*] \itemsep.5pt
\item We classify a galaxy as an elliptical galaxy when it is a spheroidal dominant system without any distinct disk or clumpy features. Elliptical galaxies are divided into subclasses depending on the projected ellipticity defined as $\epsilon$ = 1$-$IsoB$_{\rm r}$/IsoA$_{\rm r}$. These subclasses are only derived from the projected shape and thus may not be highly indicative of the true shapes \citep{kim07}.

\item Lenticular galaxies are bulge dominant but also have a disk feature. With the exception of the presence of a disk, the properties of lenticular galaxies are similar to those of elliptical galaxies, making it hard to distinguish between them \citep{van09}. However lenticular galaxies can be distinguished by a less concentrated and extended light distributions compared to elliptical galaxies.

\item The most distinct feature of spiral galaxies is the presence of spiral arms. Among the spiral galaxies, early-type spiral galaxies such as Sa or SBa have tightly wound spiral arms and a relatively large bulge. From S(B)a to S(B)d, spiral arms are wound less tightly and the bulge becomes less dominant. Furthermore, the presence or absence of a bar is also an important characteristics of spiral galaxies. By using the gray scale images with sharper contrast, we try to minimize the bias caused by weak bars, which are difficult to identify due to their low luminosity and short length \citep{oh12,san13}.

\item Irregular galaxies have a wide variety of forms. In general, they show weak spiral arm features and a very small bulge (or maybe bulgeless), and are rather asymmetric.

\item Galaxies in the unknown class are the most difficult to define. The majority of them may be lenticular galaxies which tend to be confused with elliptical galaxies, as mentioned above. Many small (though brighter than our limit) galaxies, without any distinct feature, were difficult to classify and thus placed into the unknown category. Very faint galaxies, with few striking features, and also galaxies that are disturbed by tidal interactions or mergers, are also included in the unknown category.
\end{enumerate}

% --------------------------------------------------------------------------------

% Figure - fraction
\begin{figure}
\begin{center}
\includegraphics[width=0.5\textwidth]{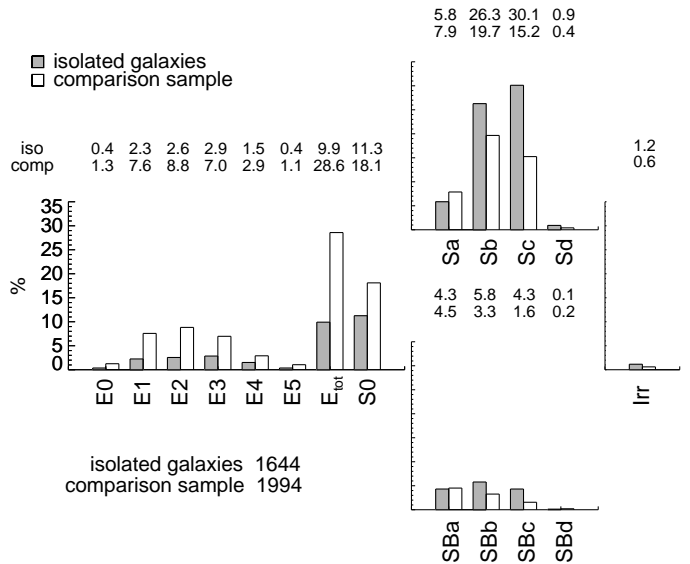}
\caption{Percentage of galaxies of each morphological type presented in the form of the Hubble tuning fork diagram. Grey bars represent fractions of galaxies from the isolated sample, and white bars show those of the comparison galaxies. Number above each bar indicate the percentage of galaxies for given morphology.}
\label{fig:fraction}
\end{center}
\end{figure}

% Figure - Concentration index
\begin{figure}
\begin{center}
\includegraphics[width=0.5\textwidth]{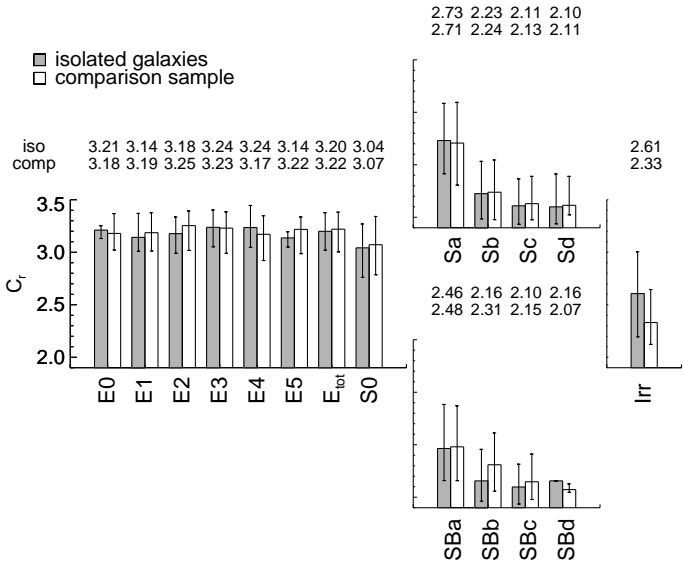}
\caption{Concentration index. Bars and associated error bars indicate median values and the standard deviations. The median values are also printed in the figures. All values in this and subsequent figures are presented in Table~\ref{tab:catalogone} --~\ref{tab:catalogfour}.}
\label{fig:cr}
\end{center}
\end{figure}

\subsubsection{Morphology Distribution}
\label{sec:mor_distribution}
By excluding unknown galaxies from each sample, our sample has 1,644 isolated galaxies and 1,994 comparison sample members. The similarity in size of the two samples makes our statistical analysis straightforward. Figure~\ref{fig:fraction} shows the distribution of morphology fraction along the Hubble sequence. The fraction of isolated galaxies and comparison sample galaxies are presented in gray and white bars, respectively. E$_{\rm tot}$ indicates the aggregates of subclasses of elliptical galaxies (E0--E5). Early-type galaxies are more common in the comparison samples. This is in agreement with the well-known morphology-density relation that high density environments are rich in elliptical galaxies, while low density regions are spiral-rich \citep{dre80}.

Specifically, elliptical galaxies are more abundant in the comparison samples than in isolated galaxies by a factor of 3 (28.6\% of comparison sample and 9.9\% of isolated galaxies). Since the work of \citet{too72}, it is generally accepted that a merger of two disk galaxies can result in the formation of an elliptical galaxy \citep{arp66,bar88,her92,naa99}. Therefore, assuming the merger rate is proportional to the density of the environment, it is natural that the fraction of elliptical galaxies is higher in the comparison sample. We must note, however, that the comparison sample galaxies are in diverse environments spanning from a small galaxy group to a large galaxy cluster. In a group environment, fly-by tidal interactions and mergers may occur frequently \citep{ost80,mak97,per09}. On the other hand, in the cluster environment, mergers are unlikely to occur due to the fast relative motion of galaxies within the deep gravitational potential well of the cluster \citep{bin87}. Interestingly, deep optical observations show the frequency of the merger features in clusters is almost comparable to that in the field \citep{she12}. This contradiction seems to be explained by the merger relics scenario suggested by \citet{yi13}. In this scenario, galaxies with merger features flow into the cluster, having experienced their merger events outside the cluster environment. While complexities exist, it is probably natural to assume that today's dense regions were assembled by smaller density peaks in the earlier universe where mergers were frequent.

In both samples, slightly oval shaped galaxies such as E2 or E3 are greatest in abundance and this abundance decreases as galaxies become rounder or flatter, confirming the result of \citet{oh13}. According to \citet{kim07}, luminous elliptical galaxies are likely to be triaxial and less luminous elliptical galaxies are usually oblate. However, we cannot find any prominent distinction between the two samples. It is probably because our classification scheme doesn't reflect their intrinsic shapes, but merely the projected ones.

% Figure - FracDeV
\begin{figure}
\centering
\includegraphics[width=0.5\textwidth]{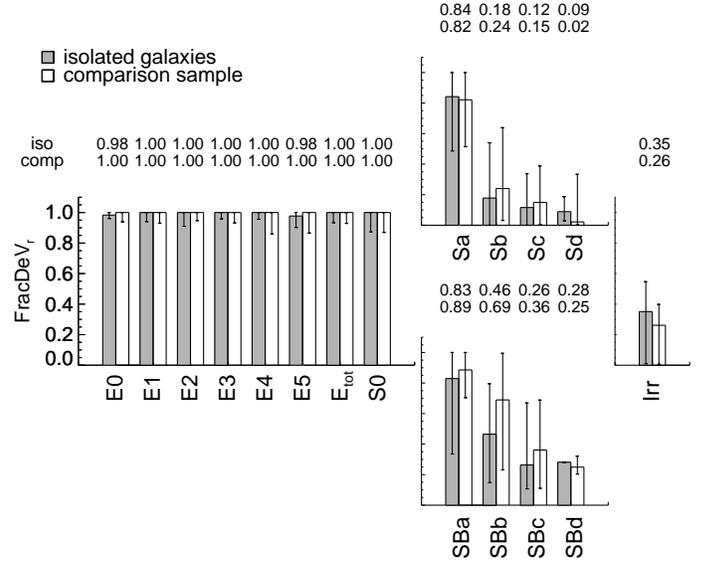}
\caption{FracDeV of each morphological type. The format is the same as that of Figure~\ref{fig:cr}.}
\label{fig:fracdev}
\end{figure}

Sa galaxies are more frequently found in the comparison sample, but Sc galaxies are more abundant in isolated galaxies (7.9\%, 19.7\%, 15.2\% in comparison samples and 5.8\%, 26.3\%, 30.1\% in isolated galaxies for Sa, Sb and Sc, respectively). This trend is the same in barred spiral galaxies. Furthermore, for the isolated galaxy sample, there is a steady increase in number as we move from Sa to Sc galaxies, which is consistent with other studies \citep{her08,kar10}. On the other hand, comparison sample galaxies do not show such an increasing trend from Sa to Sc type. This implies that the evolutionary histories of spiral galaxies in the isolated systems and relatively denser regions may be different.

% Figure - Morphology
\begin{figure*}
\begin{center}
\includegraphics[width=0.85\textwidth, angle=270]{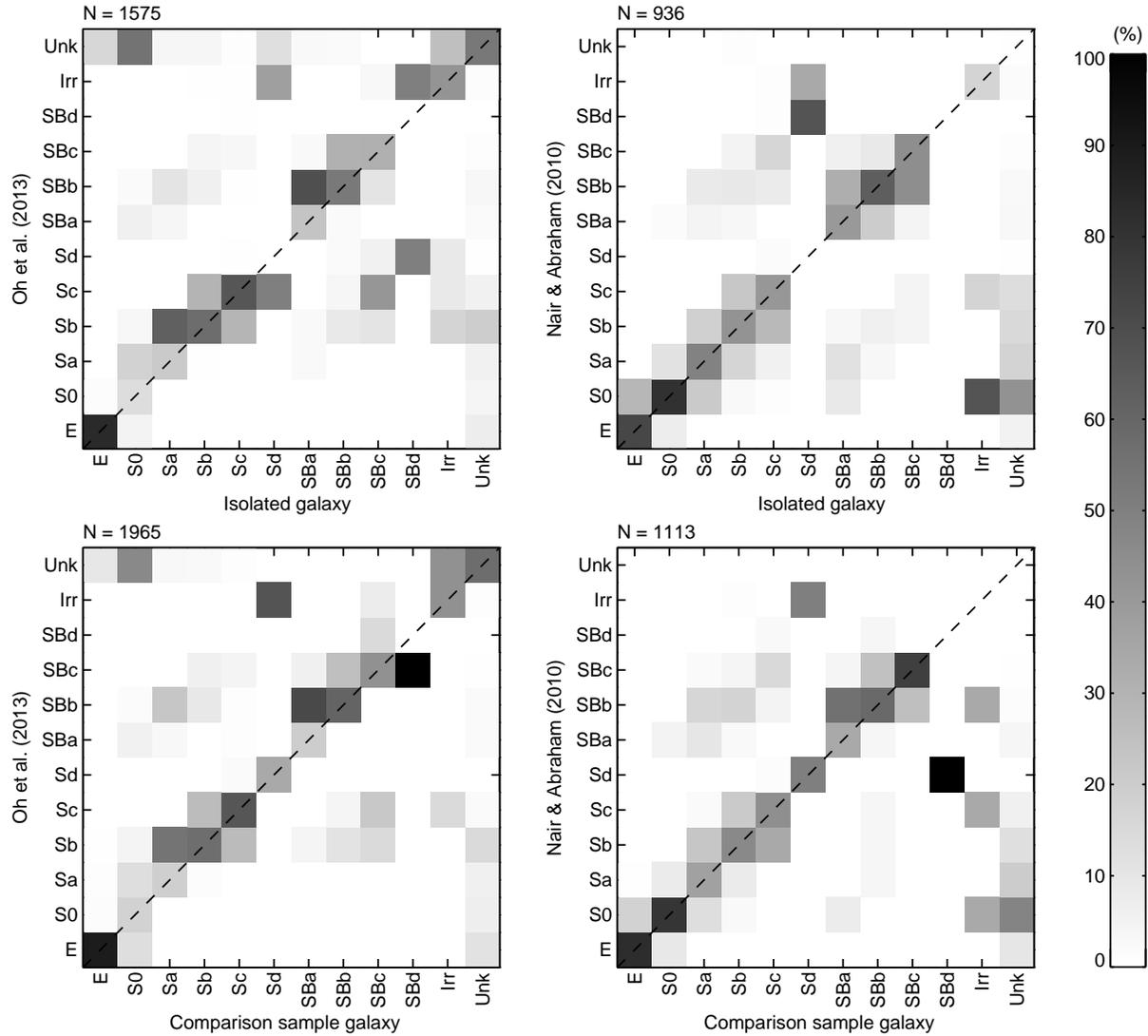}
\caption{Comparison of morphology classifications. We compared our classifications with \citet{oh13} (left panels) and \citet{nai10a} (right panels). Upper panels are for isolated galaxies and lower panels for comparison sample galaxies.}
\label{fig:compare}
\end{center}
\end{figure*}

The mechanism for bar formation is debatable. The bar structures can be formed through either a secular evolution \citep{too64,jog84} or during a tidal interaction with another galaxy \citep{nog87,elm90,elm91}. Thus, it is interesting to compare the bar fractions\footnote{$f_{\rm bar} =$ $N_{\rm barred} \over  N_{\rm unbarred}+N_{\rm barred}$} of the two samples in different density regions. \citet{mar09} reported that the bar fraction in the clusters is comparable to that of field galaxies at lower redshift, whereas \citet{ver07a} found no environmental preference. Furthermore, \citet{lee12} found the fraction of barred galaxies are not influenced by the nearest neighbor galaxy if the separation between galaxies is larger than 0.1 times the virial radius of the neighbor. On the other hand, \citet{var04} claims that bars are more frequently found in perturbed late-type galaxies than in isolated ones by a factor of 2. We found that 18.7\% and 18.1\% of spiral galaxies are barred spiral galaxies in the isolated system and comparison sample, respectively\footnote{Our bar fraction is a bit lower than those in some previous studies. In those studies, however, an inclination cut is used to avoid the bar obscuration by the galactic disk. When we apply IsoB$_{\rm r}$/IsoA$_{\rm r}>$ 0.7 as used in \citet{oh12} for example, the bar fraction increases up to 23.8\% and 25.1\% for isolated and comparison sample galaxies, respectively. These values are close to the bar fraction found in other studies which use a visual inspection for the bar identification \citep[e.g.,][]{mas10,nai10b,lee12,oh12}. We acknowledge this systematic uncertainty, and we thus focus only on the difference between isolated and comparison samples.}. This result indicates that there is no significant difference in the bar fraction as a function of density in our sample. While we admit that our environmental criterion is not comprehensive, one may be tempted to interpret this result as evidence against the environmental scenario for the bar formation.

% --------------------------------------------------------------------------------

\subsubsection{Parameters Depending on Morphology}
The concentration index is widely regarded as a tool for morphology classification \citep{doi93,abr94,abr96,shi01}. Since we classified the morphologies of galaxies using a visual inspection, the concentration index can be used as a validity check for our classification. There are several distinct definitions of concentration index, but here we use the ratio between PetroR90 and PetroR50 of the $r$-band petrosian magnitude, i.e. C$_{\rm r}$ $\equiv$ PetroR90/PetroR50.  Figure~\ref{fig:cr} shows C$_{\rm r}$ values on the Hubble tuning fork. The bar indicates the median value and the error bar is the 1$\sigma$ dispersion. Elliptical galaxies usually have a concentration index greater than 3 with small dispersion, while late-type galaxies typically have smaller C$_{\rm r}$. For all Hubble-types, there is no significant difference in concentration index between the isolated and comparison galaxies.

A similar test can be performed on $\fracDeV$ in $r$-band which is a photometric parameter provided by SDSS. This value indicates the fraction of flux which the de Vaucouleurs profile accounts for. Therefore, an elliptical galaxy tends to have $\fracDeV$ close to unity, whereas a spiral galaxy has a lower $\fracDeV$. In Figure~\ref{fig:fracdev}, the median values of $\fracDeV$ of elliptical galaxies are close to 1, regardless of the subclasses. For late-type galaxies, $\fracDeV$ decreases from S(B)a to S(B)d, as expected. And also, like  C$_{\rm r}$, we cannot find any prominent difference in $\fracDeV$ between our different density samples.

% --------------------------------------------------------------------------------

\subsubsection{Comparison with Previous Works}
\label{sec:morph_compare}
In this section, we compared our morphological classification result with those of other studies. \citet{oh13} classified the morphology of 10,233 galaxies for the SDSS DR7 database using a visual inspection based on the Hubble classification scheme. Their redshift range ($0.033 < z < 0.044$) is similar to ours, and thus there is some overlap with our sample. Among isolated galaxies (2,660) and comparison sample galaxies (3,409), 1,575 and 1,965 galaxies are matched with the galaxies in \citet{oh13}, respectively. Similarly, \citet{nai10a} performed visual inspections for 14,034 galaxies in the SDSS DR4 following their own classification scheme which is similar to RC3 classifications under the consideration of fine structures like rings, lenses and shells. They use galaxies with an apparent limit of g $< 16$ mag, and in the redshift range of $0.01 < z < 0.1$ which covers our sample's redshift range. As a result, 936 isolated galaxies and 1,113 comparison sample galaxies of our galaxies overlap with their sample.

Figure~\ref{fig:compare} shows the result of morphological comparison for isolated galaxies (upper panels) and comparison sample galaxies (lower panels). The percentage represented by gray scale shading is calculated by counting the number of galaxies that match the morphological type in the other studies. On the whole, our classification is in good agreement with others. For all morphologies except the unknown type, $\sim$85\% of galaxies in each panel are near the one-to-one line. There are a number of galaxies classified as S0 galaxies in \citet{nai10a} but as unknown in our scheme, meaning that we were more conservative in our classification of S0 galaxies (see Section~\ref{sec:morph}) than \citet{nai10a}.

% --------------------------------------------------------------------------------

% Figure - CMR
\begin{figure}
\centering
\includegraphics[width=0.5\textwidth]{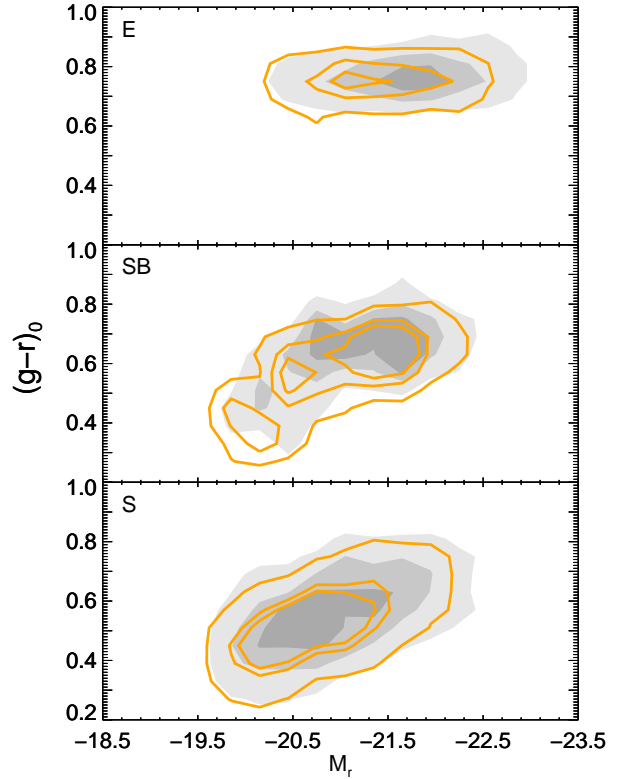}
\caption{Color-magnitude diagram for elliptical (upper panel), barred spiral (middle panel) and unbarred spiral galaxies (lower panel). Orange contours indicate isolated galaxies and shaded regions show the comparison sample. From the inside out, contours at 0.5$\sigma$, 1.0$\sigma$, and 2.0$\sigma$ are drawn.}
\label{fig:cmr}
\end{figure}

\subsection{Color-Magnitude Diagram}
\label{sec:cmr}
Depending on the morphological type, the galaxies tend to reside in different regions on the color-magnitude diagram. The magnitude and color are Galactic extinction-corrected with the data from the SDSS pipeline and $k$-corrected using $\kcorrect$ v4.2 package \citep{bla07}. Early-type galaxies lie on a well-defined red-sequence, whose tilt shows that massive early-type galaxies are redder than less massive ones \citep{san78,bow92,dri06}. In Figure~\ref{fig:cmr}, for both isolated galaxies (solid line) and comparison sample galaxies (shade region), there are tight color-magnitude relations (CMR) of elliptical galaxies. Elliptical galaxies in the isolated system are fainter ($M_{\rm r} = -21.36$), on the whole, than those in high density environments ($M_{\rm r} = -21.73$) with an offset in median $M_{\rm r}$ by $\sim$0.4. This result can be understood within the hierarchical merger scenario if galaxies in denser regions have more opportunity to become massive through mass accretion and mergers \citep[e.g.,][]{bau96,kau96,kau98,col00,deL06,lee13,tar13}.

Furthermore, for the galaxies with $M_{\rm r} < -21.5$, \citet{sch07b} found that the NUV-r color of early-type galaxies in low-density environments extends further into the blue than that of early-type galaxies in high-density environments, indicating the presence of residual star formation \citep{yi05,kav07,jeo09}. However, we cannot find a prominent vertical offset in color, probably because the (g--r)$_0$ color is not very sensitive to the residual star formation \citep{sch07b}.

Unbarred spiral galaxies prefer to be located in the blue cloud as opposed to in the red sequence \citep{str01,bel03,bla03,hog03}. Unbarred spiral galaxies for both isolated and comparison sample are in the blue cloud, and the overall distributions are similar (the lower panel of Figure 6). The offset in median $M_{\rm r}$ ($\sim$0.15) between isolated and comparison sample galaxies is less significant than that found in elliptical galaxies. Also, the difference in median (g-r)$_0$ color between the two samples is only $\sim$0.03 which may result from the different median $M_{\rm r}$. This value is consistent with that found in the literature \citep{hog04,var04}.

In contrast to unbarred spirals, barred spiral galaxies show a bimodality with two peaks; a luminous and red peak, and a faint and blue peak (middle panel of Figure~\ref{fig:cmr}). This is consistent with other studies (\citealt{bar08,nai10b,oh12}; \citealt{oh13}), although the strength of the second peak can vary depending on the methods for identifying a weak bar or the sample selection criteria \citep{mar09,mas11,lee12,san13}. The bimodality appears in both the isolated and comparison sample galaxies. There is no obvious difference between the contours of the two samples, nor in the bar fraction (see Section~\ref{sec:mor_distribution}).

% --------------------------------------------------------------------------------

% Figure - Stellar mass
\begin{figure}
\begin{center}
\includegraphics[width=0.5\textwidth]{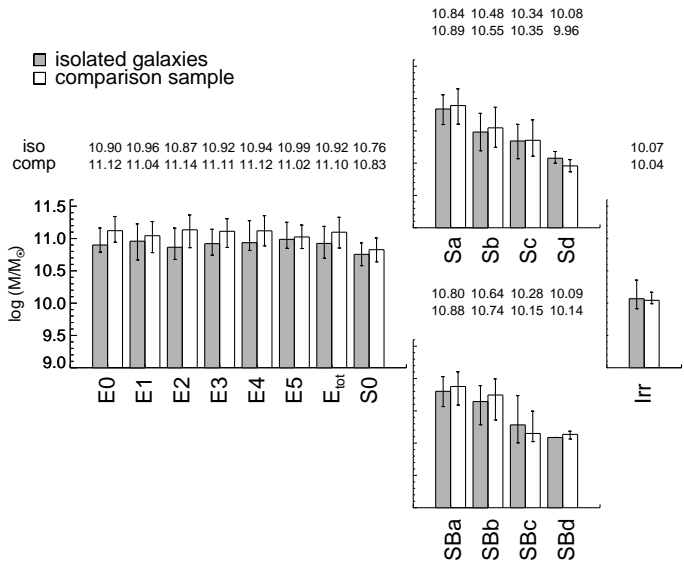}
\caption{Stellar mass. The format is the same as that of Figure~\ref{fig:cr}.}
\label{fig:mass}
\end{center}
\end{figure}

\subsection{Stellar Mass}
\label{sec:mass}
The stellar mass is calculated using the formula in \citet{bel03} with the $k$-corrected g--r color and $r$-band magnitude. The stellar mass along the Hubble tuning fork is shown in Figure~\ref{fig:mass}. The median value of the stellar mass of early-type galaxies is higher than that of late-type galaxies in both isolated and comparison sample as expected. Among elliptical galaxies, there is no big difference among the subclasses, because the subclass is not revealing the true geometry due to projection effects \citep{kim07}. In late-type galaxies, the stellar mass becomes smaller as the galaxies tend to be less bulge dominated for both barred and unbarred spiral galaxies \citep{oh13}, regardless of the surrounding environmental density.

The median stellar mass of elliptical galaxies (E$_{\rm tot}$) in the comparison sample ($10^{11.10}\ M_\odot$) is about 50\% larger than that of isolated elliptical galaxies ($10^{10.92}\ M_\odot$). This can be understood as a result of difference in the merger and accretion history as mentioned previously. Meanwhile, for the stellar mass of late-type galaxies, there are less prominent distinctions (10--20\%) between the isolated galaxies and comparison sample galaxies along Hubble types. This may imply that the majority of mass of late-type galaxies come from the in-situ star formation accompanied with a minor contribution from the merger \citep{tot92,mal06,bou07}, regardless of the environment density. The results are consistent when we instead use the stellar mass from the MPA-JHU catalog.

% --------------------------------------------------------------------------------

% Figure - sigma
\begin{figure}
\begin{center}
\includegraphics[width=0.5\textwidth]{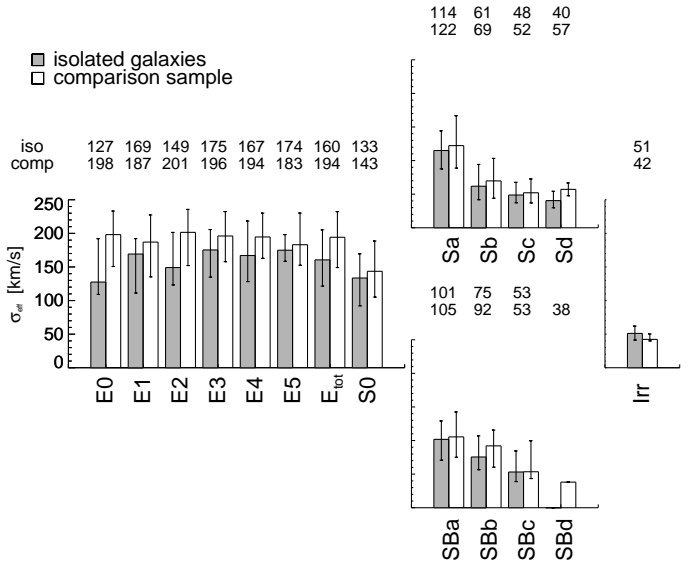}
\caption{Velocity Dispersion. The format is the same as that of Figure~\ref{fig:cr}.}
\label{fig:sigma}
\end{center}
\end{figure}

\subsection{Velocity dispersion}
We use the velocity dispersion provided by the OSSY catalog \citep{oh11}, that were measured using $\gandalf$ based on the improved spectral fits. Since the SDSS performed the spectroscopy using fibers with 3$\arcsec$ diameter, we use the effective radius of galaxies \citep{gra05} in order to obtain the effective velocity dispersion \citep{cap06}, corrected for the size of the fiber. We chose only galaxies with $10<\sigma_{\rm eff}<400$ \kms and $error(\sigma_{\rm eff})$/$\sigma_{\rm eff}< 0.5$. As a result, we have 1,386 (84\%) and 1,816 (91\%) galaxies for the isolated and comparison sample, respectively. Most of the early-type galaxies (99.9\%) and S(B)a and S(B)b (92\%) satisfy this condition, but plenty of S(B)c and S(B)d (68\% remain) are removed due to their intrinsically low velocity dispersions making it difficult to measure with high precision.

Whether in the isolated or comparison sample, elliptical galaxies have larger velocity dispersions than late-type galaxies (Figure~\ref{fig:sigma}), which is consistent with previous studies \citep{veg01,agu09}. For elliptical galaxies, the median value of the velocity dispersions of comparison sample galaxies (194$^{+38}_{-45}$ \kms) is higher than that of isolated galaxies (160$^{+44}_{-39}$ \kms). However, the velocity dispersions of spiral galaxies shows only small differences between isolated and comparison sample galaxies. This trend is similar to that of the stellar mass because the velocity dispersion is related to the dynamical mass, which may roughly trace the stellar mass.

% --------------------------------------------------------------------------------

\section{Absorption-line Statistics}
\label{sec:absorption}
Absorption lines of galaxies are essential tools to study their stellar populatons, which in turn provide information on the history of a galaxy's formation and evolution. In this study, we use measurements of absorption lines given by the OSSY catalog \citep{oh11} to study stellar populations. The OSSY catalog, based on the Lick/IDS system resolution, provides the strength of absorption lines which are measured taking into consideration for the emission line subtraction using the $\gandalf$ fit tool. Accurate subtraction of emission lines is important, because emission lines near absorption lines are likely to hinder determining the pseudocontinuum properly. We apply a signal-to-noise cut ($S/N > 3$) and quality-assessing parameter (N$\sigma<2$) to make the absorption line strength more reliable. As a result, 98.4\% and 98.8\% of galaxies for the isolated and the comparison sample remained, respectively. As the line measurements only represent the spectroscopic properties of the central part of galaxies due to the size of SDSS fiber (3\arcsec), they cannot be compared directly with the photometric characteristics. However using both sets of data can help to understand the overall properties of the galaxies.

% --------------------------------------------------------------------------------

% Figure - H beta
\begin{figure}
\begin{center}
\includegraphics[width=0.5\textwidth]{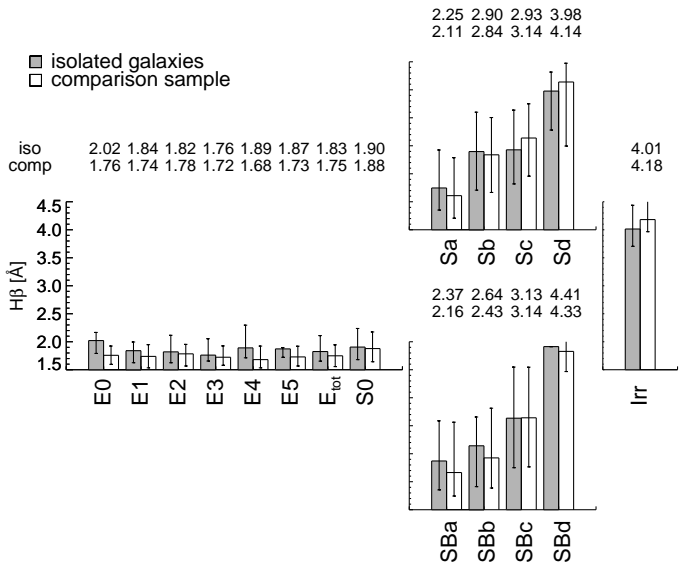}
\caption{H$\beta$ absorption line strengths. The format is the same as that of Figure~\ref{fig:cr}.}
\label{fig:hb}
\end{center}
\end{figure}

\subsection{H$\beta$}
\label{sec:sec_hb}
The strength of absorption lines are generally affected by temperature and metallicity. Among the absorption lines, Balmer absorption line indices are sensitive to the temperature, and strongest around an effective temperature of 10,000 K (A-type stars). Therefore, an old stellar population which has a relatively low temperature shows a weak Balmer absorption line in general \citep{wor94,tra00}. For the analysis of Balmer absorption line, we use the H$\beta$ absorption line.

Equivalent width (EW) values of the H$\beta$ absorption line are shown in Figure~\ref{fig:hb}. The H$\beta$ absorption line strength of early-type galaxies is lower than that of late-type galaxies \citep{tan98,ber06,kun06,gan07}, indicating that the stellar component of early-type galaxies is older, as expected. Also, in late-type galaxies, the median EW of H$\beta$ absorption line becomes larger along the Hubble sequence for both isolated and comparison sample galaxies. This is consistent with the idea that more late-type spiral galaxies are more gas rich, with a higher star formation rate (SFR), and thus relatively younger.

Although the median EW of the H$\beta$ absorption line index in isolated elliptical galaxies (1.83$^{+0.28}_{-0.17}$ \AA) is slightly larger than that of elliptical galaxies in the comparison sample (1.75$^{+0.20}_{-0.19}$ \AA) and this trend happens for all subclasses of elliptical galaxies, the differences are not notable. For spiral galaxies, they don't show any prominent difference. Since the strength of the H$\beta$ absorption line is also affected by metallicity, a more detailed analysis will be performed in Section~\ref{sec:sec_pop}.

% --------------------------------------------------------------------------------

% Figure - Fe5270
\begin{figure}
\begin{center}
\includegraphics[width=0.5\textwidth]{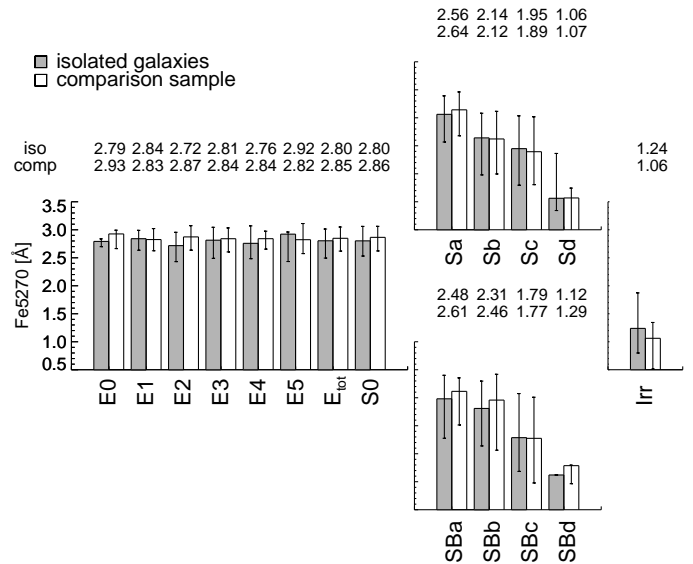}
\caption{Fe5270 absorption line strengths. The format is the same as that of Figure~\ref{fig:cr}.}
\label{fig:fe5270}
\end{center}
\end{figure}

\subsection{Fe5270}
\label{sec:sec_fe5270}
According to \citet{lar74}, massive galaxies with deep gravitational potential wells can better maintain their interstellar media against ejection by SNe explosions, leading to more efficient chemical recycling. Therefore in this scenario, massive galaxies are expected to be more metal rich, as observed in the well-known mass--metallicity relation \citep[e.g.,][]{tre04}. To study the metallicity, we chose the Fe5270 absorption line index as a representative metal line which is a typical Fe indicator used in the stellar population model \citep{tho03}.

Figure~\ref{fig:fe5270} shows, for both isolated and comparison samples, early-type galaxies have larger EW values of the Fe5270 absorption line than late-type galaxies do. This can be inferred from the fact that early-type galaxies are more massive than late-type galaxies (Figure~\ref{fig:mass}) resulting in higher metallicity. Furthermore, late-type galaxies have descending trends of the Fe5270 line strength from S(B)a to S(B)d, regardless of the density environment, and this can also be understood in terms of the mass-metallicity relation. Like H$\beta$, there is a small difference of Fe5270 line strength between isolated and comparison sample galaxies, for both elliptical and spiral galaxies, but it is not significant enough to compare their metallicity without consideration of the degeneracy of age and metallicity.

% --------------------------------------------------------------------------------

% Figure - Mgb
\begin{figure}
\begin{center}
\includegraphics[width=0.5\textwidth]{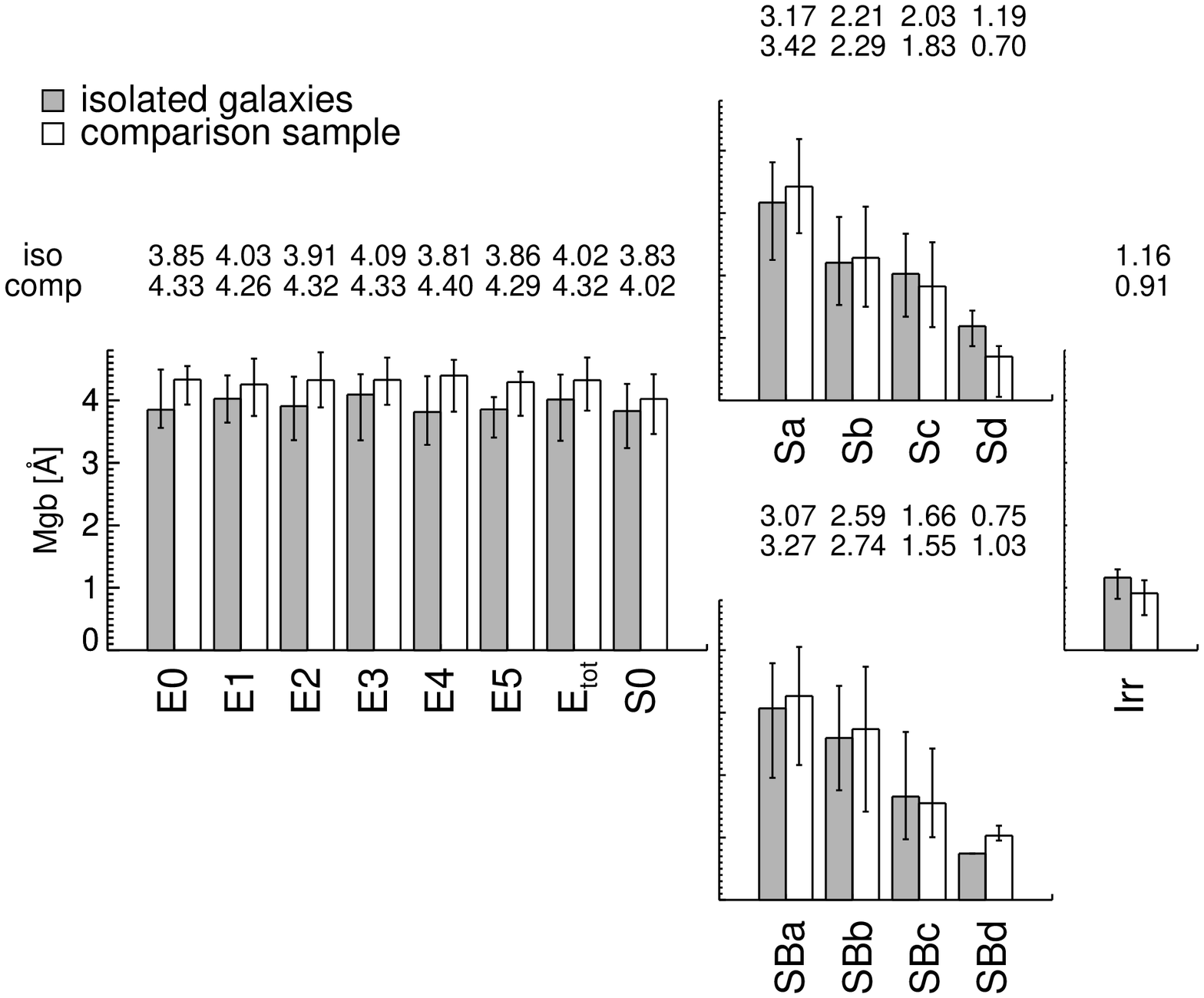}
\caption{Mgb absorption line strenghs. The format is the same as that of Figure~\ref{fig:cr}.}
\label{fig:mgb}
\end{center}
\end{figure}

\subsection{Mgb}
\label{sec:sec_mgb}
Type II supernovae are thought to occur shortly after the onset of star formation. As a result, they quickly spread $\alpha$-elements, their main product, into the interstellar medium. Meanwhile, Type Ia supernovae whose generation is delayed, but relatively continuous throughout the period of star formation, supply predominantly Fe-peak elements \citep{tin79,nom84,woo95,thi96}. Thus, by comparing the abundance of $\alpha$-elements and Fe, [$\alpha$/Fe], we can trace the star formation history of galaxies \citep{gre83,mat86,mat87,pag95,mcW97}.

The Mgb absorption line is widely used to represent the abundance of $\alpha$-elements. We can see the Mgb absorption line strength of early-type galaxies is larger than that of the late-type galaxies. Also among late-type galaxies, the strength decreases from S(B)a to S(B)d for both isolated and comparison sample galaxies (Figure~\ref{fig:mgb}). This result implies that the early-type galaxies experience a relatively intensive star formation activity, while the late-type galaxies have an extended star formation history.

For elliptical galaxies, the median EW of Mgb of isolated galaxies (4.02$^{+0.40}_{-0.66}$ \AA) is a bit smaller than that of comparison galaxies (4.32$^{+0.36}_{-0.49}$ \AA), but the difference is not large. For spiral galaxies, the difference is even less significant.

% --------------------------------------------------------------------------------

% Figure - MgFe_hb
\begin{figure}
\begin{center}
\includegraphics[width=0.34\textwidth, angle = 270]{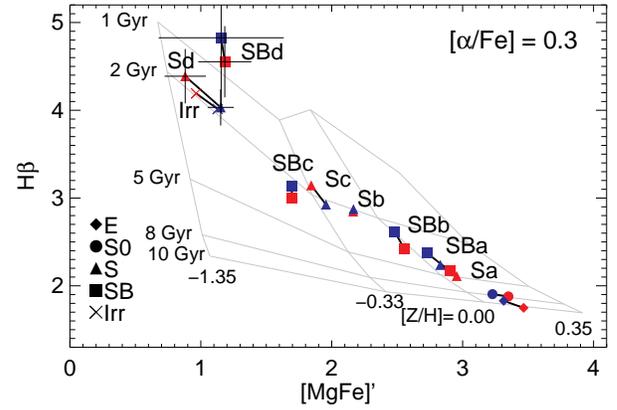}
\caption{[MgFe]$\arcmin$--H$\beta$ plane for [$\alpha$/Fe] = 0.3. The model grids are from TMJ model \citep{tho11}. For representative values, median values for [MgFe]$\arcmin$ and H$\beta$ are used for each morphology. Isolated galaxies and comparison ones are presented in blue and red color, respectively. Here, error bars representing observation errors are drawn only when error bars are greater than the symbol size. Galaxies with same morphology in different samples are linked by a solid line.}
\label{fig:mgfe-hb}
\end{center}
\end{figure}

% Figure - Mgb_Fe
\begin{figure}
\begin{center}
\includegraphics[width=0.555\textwidth, angle = 270]{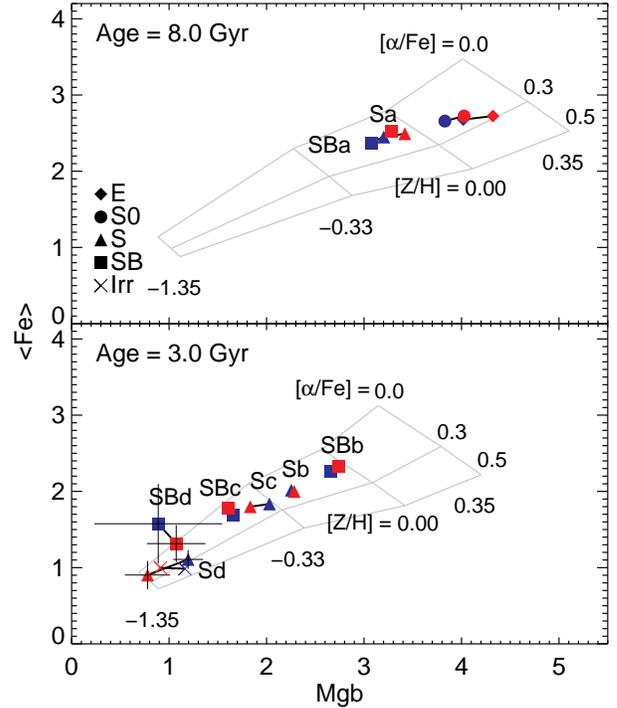}
\caption{Mgb--$<$Fe$>$ plane for 8 Gyr (upper panel) and 3 Gyr (lower panel). The notation for symbols and error bars is same as that in Figure~\ref{fig:mgfe-hb}.}
\label{fig:mgb-fe}
\end{center}
\end{figure}

% Figure - Model_age_50_err_cut
\begin{figure}
\begin{center}
\includegraphics[width=0.5\textwidth, angle = 270]{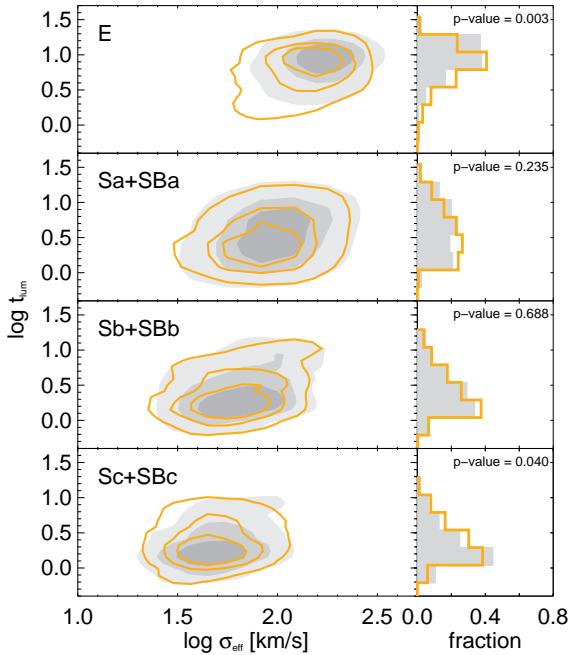}
\caption{Left panels show contours of effective velocity dispersions and luminosity-weighted ages in Gyr for elliptical, Sa-type, Sb-type and Sc-type spiral galaxies. Orange contours and shade regions indicate isolated galaxies and comparison ones, respectively. Right panels are for the histogram of luminosity-weighted ages with p-value from the K-S test. See the text for details.}
\label{fig:age}
\end{center}
\end{figure}

% --------------------------------------------------------------------------------

\subsection{Comparison with Stellar Population Models}
\label{sec:sec_pop}
As we discussed above, absorption line strengths are affected by both age and metallicity, the so called `age--metallicity degeneracy' \citep{wor94}. Thus, it is important to consider the absorption line indices such as H$\beta$, Fe5270 and Mgb simultaneously, based on the stellar population model, in order to try and break the degeneracy. In this section, we are going to discuss the stellar age, metallicity and [$\alpha$/Fe] derived from the stellar population model devised by \citet{tho11} (hereafter the TMJ model).

The stellar population models are constructed based on an evolutionary population synthesis model with calibrations (\citealt{wor94,tho03}; \citealt{tho11}). We used the TMJ model which covers a wider range of [$\alpha$/Fe] values than the model of \citet{tho03} (TMB model) used in \citet{oh13}. We should bear in mind that this method provides the ``luminosity-weighted'' age and metallicity, and thus it is biased towards the younger, brighter stellar components.

To break the degeneracy between age and metallicity, the model grid of the [MgFe]$\arcmin$--H$\beta$ plane was used. We adopted the index [MgFe]$\arcmin$ which is less sensitive to [$\alpha$/Fe] and therefore a good tracer of metallicity \citep{tho03}. We fixed [$\alpha$/Fe] as 0.3 for simplicity, but this does not affect the results significantly because neither H$\beta$ \citep{tho04,tho06} nor [MgFe]$\arcmin$ is responsive to [$\alpha$/Fe]. The median values of the luminosity-weighted age and metallicity for each morphology are shown in Figure~\ref{fig:mgfe-hb}. We did not separate elliptical galaxies into subclasses. The blue symbols indicate isolated galaxies and the red ones are comparison sample galaxies. Along the Hubble sequence, the median age and metallicity decrease for both isolated and comparison sample galaxies. There is a trend that isolated galaxies are slightly younger and more metal poor based on median values than comparison sample galaxies for almost all morphologies, but the differences are not significant (Figure~\ref{fig:mgfe-hb}).
 
We can also investigate the properties of stellar populations using the Mgb--$<$Fe$>$ plane for fixed ages. Here, $<$Fe$>$ is defined as a mean value of Fe5270 and Fe5335 absorption lines, (Fe5270+Fe5335)/2 \citep{gor90}. For a representative age, we chose the age 8 Gyr for early-type galaxies including S(B)a and 3 Gyr for late-type galaxies (Figure~\ref{fig:mgb-fe}). As with the trends of the median age and metallicity derived from the [MgFe]$\arcmin$--H$\beta$ plane, the median [$\alpha$/Fe] decreases along the Hubble sequence. However, the median values of [$\alpha$/Fe] of isolated galaxies do not seem to be significantly different from those of comparison sample galaxies. Metallicities derived from the Mgb--$<$Fe$>$ plane also show a weak offset between the two samples as seen in Figure~\ref{fig:mgfe-hb}.

For a quantitative analysis, we derived the luminosity-weighted age, [Z/H] and [$\alpha$/Fe] for each galaxy using an iterative method. This method is based on a circular logic; we firstly estimate the age for fixed [$\alpha$/Fe] on the [MgFe]$\arcmin$--H$\beta$ plane, and then using the estimated age, derive the [$\alpha$/Fe] on the Mgb--$<$Fe$>$ plane. By iterating this process for each galaxy, the quantity converges to a single value.

% Figure -Model_z_h_50_err_cut
\begin{figure}
\begin{center}
\includegraphics[width=0.5\textwidth, angle = 270]{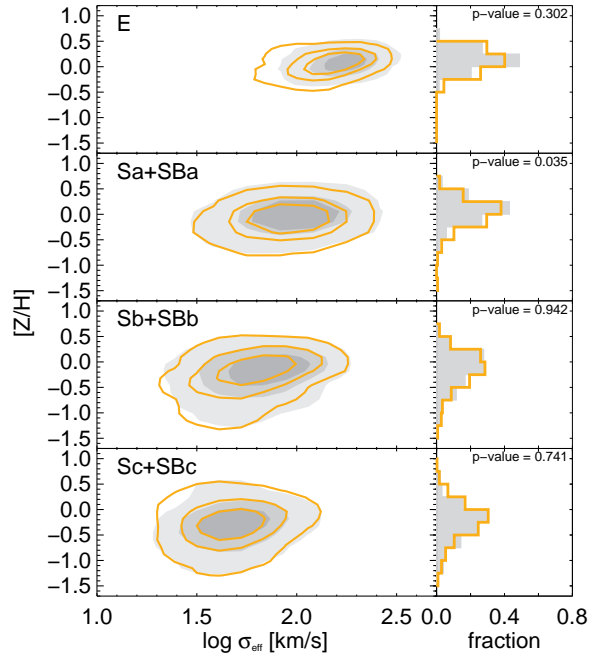}
\caption{Same as Figure~\ref{fig:age}, but for the luminosity-weighted [Z/H].}
\label{fig:z_h}
\end{center}
\end{figure}

Figure~\ref{fig:age} shows contours (left panels) for luminosity-weighted ages of isolated (orange solid line) and comparison galaxies (shade region) along the effective velocity dispersion. For a fixed velocity dispersion, the luminosity-weighted age does not differ significantly between the two samples. However, the contour of isolated elliptical galaxies is notably extended to lower velocity dispersions and younger ages than that of their comparison galaxies. Therefore, the median luminosity-weighted age (a representative age of the group) is likely to be younger in the isolated elliptical galaxies than the comparisons. This result is consistent with other studies which argue that, on average, the elliptical galaxies in low-density systems tend to be less massive and younger than those in high-density regions \citep{bau96,kau98,gov99,col00,kun02,tho10}. Other morphologies seem to be almost identical at all contour levels.

Right panels in Figure~\ref{fig:age} show the fraction of galaxies corresponding to each age bin. The Kolmogorov-Smirnov (K-S) test is also performed to address the similarity of distributions of isolated and comparison sample galaxies. For elliptical galaxies, the distributions of the luminosity-weighted age of the two samples seem slightly different. If we take the significance level as 0.05 (or 5\%) which is typically used, the small p-value of elliptical galaxies (0.003) suggests the younger median luminosity-weighted age of isolated galaxies (8.3 Gyr) than that of comparisons (10.0 Gyr) is statistically significant. In addition, Sc-type (Sc and SBc) spiral galaxies have the p-value (0.04) smaller than 0.05, implying there is a chance that the distributions of isolated and comparison galaxies arise from different parent samples. However, there is no obvious difference in the median luminosity-weighted age of isolated galaxies (2.0 Gyr) and comparison ones (1.9 Gyr). For Sa-type (Sa and SBa) and Sb-type (Sb and SBb) spiral galaxies, the p-values are 0.235 and 0.688, respectively and we cannot see any notable differences in ages between isolated and comparison galaxies in each histogram.

% Figure - Model_a_Fe_50_err_cut
\begin{figure}
\begin{center}
\includegraphics[width=0.5\textwidth, angle = 270]{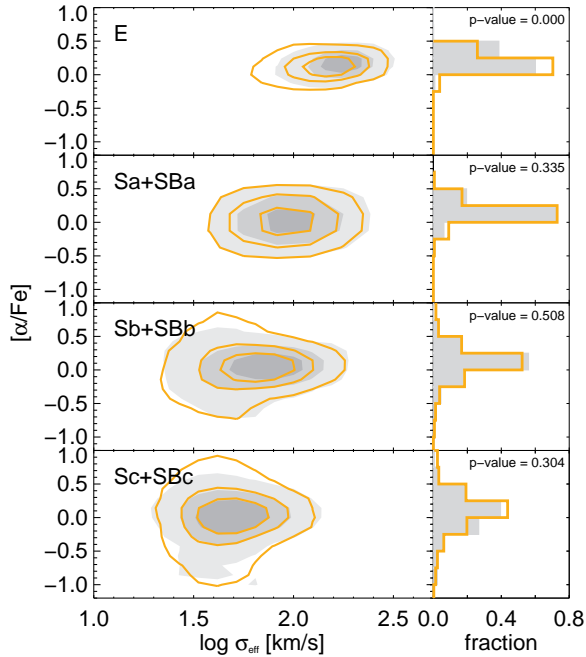}
\caption{Same as Figure~\ref{fig:age}, but for the luminosity-weighted [$\alpha$/Fe].}
\label{fig:a_fe}
\end{center}
\end{figure}

% Figure - BPT diagram
\begin{figure*}[t]
\centering
\includegraphics[width=0.85\textwidth]{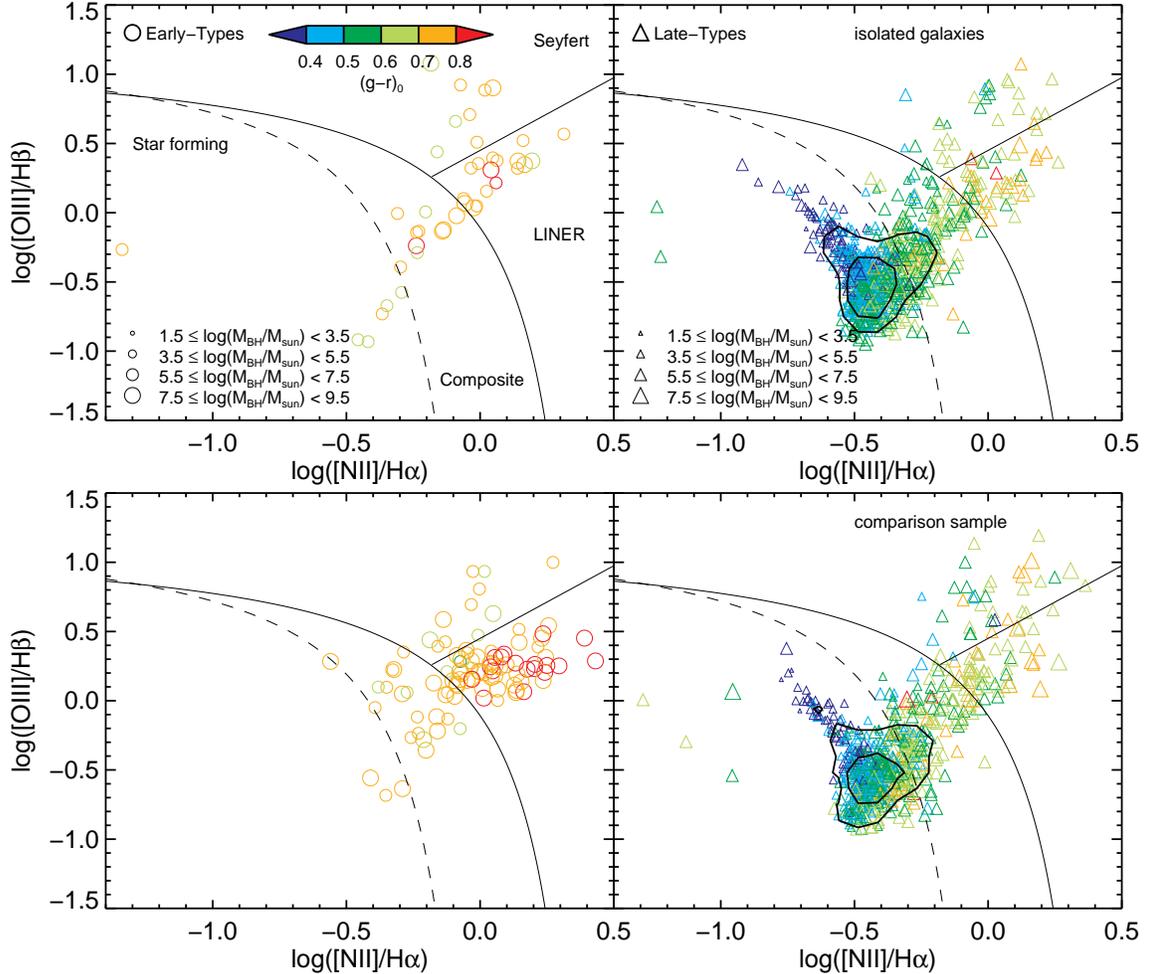}
\caption{BPT diagrams of isolated galaxies (upper panels) and comparison samples (lower panels). Galaxies with A/N $\ge$ 3 for all four emission lines are plotted. Left panels are analysis of early-type galaxies and right ones show late-type galaxies. The color and size of each symbol correspond to the (g-r)$_0$ color index and black hole mass, respectively. The dashed line is the demarcation between star-forming and composite class \citep{kau03} and solid line is the criterion for composite and AGN \citep{kew01}. The solid-straight line distinguishes Seyferts from LINERs \citep{sch07a}. In the right two panels, we draw the contours which account for 38\% (0.5$\sigma$, inner contour) and 68\% (1.0$\sigma$, outer contour) of the distribution.}
\label{fig:bpt}
\end{figure*}

The results of metallicity with the effective velocity dispersion are presented in Figure~\ref{fig:z_h}. It is difficult to discern any discrepancy between isolated galaxies and comparison samples for all morphologies in the contour diagram, except for slightly lower velocity dispersions of ellipticals in isolated galaxies. Also, histograms of the metallicity distribution for isolated and comparison galaxies seem similar for all panels. The K-S test for the metallicity distribution indicates Sa-type spiral galaxies have small p-values. But the median luminosity-weighted metallicities of isolated and comparison galaxies do not show any prominent differences.

The situation of [$\alpha$/Fe] is similar. Only elliptical galaxies show contours slightly extended to lower velocity dispersions in isolated galaxies and there is no significant difference between isolated and comparison galaxies for late-type galaxies in contour diagrams (Figure~\ref{fig:a_fe}). The p-value derived from the K-S test implies late-type galaxies are unlikely to have different [$\alpha$/Fe] distributions for those two samples. Meanwhile, the lower p-value of elliptical galaxies suggest different distributions, although there is no significant discrepancy in the median values between the isolated and comparison galaxies.

% --------------------------------------------------------------------------------

\section{Emission-Line Statistics}
\label{sec:emission}
Unlike absorption lines which indicate the cumulative star formation history, emission lines are more suitable for revealing ongoing star formation. Furthermore, active galactic nuclei (AGNs) which show strong emission lines are thought to play an important role in galaxy evolution through AGN feedback. Therefore, we now consider the properties of the emission lines. We use the flux and EW values of emission lines from the OSSY catalog.

% --------------------------------------------------------------------------------

% Figure -BPT fraction
\begin{figure}
\centering
\includegraphics[width=0.5\textwidth]{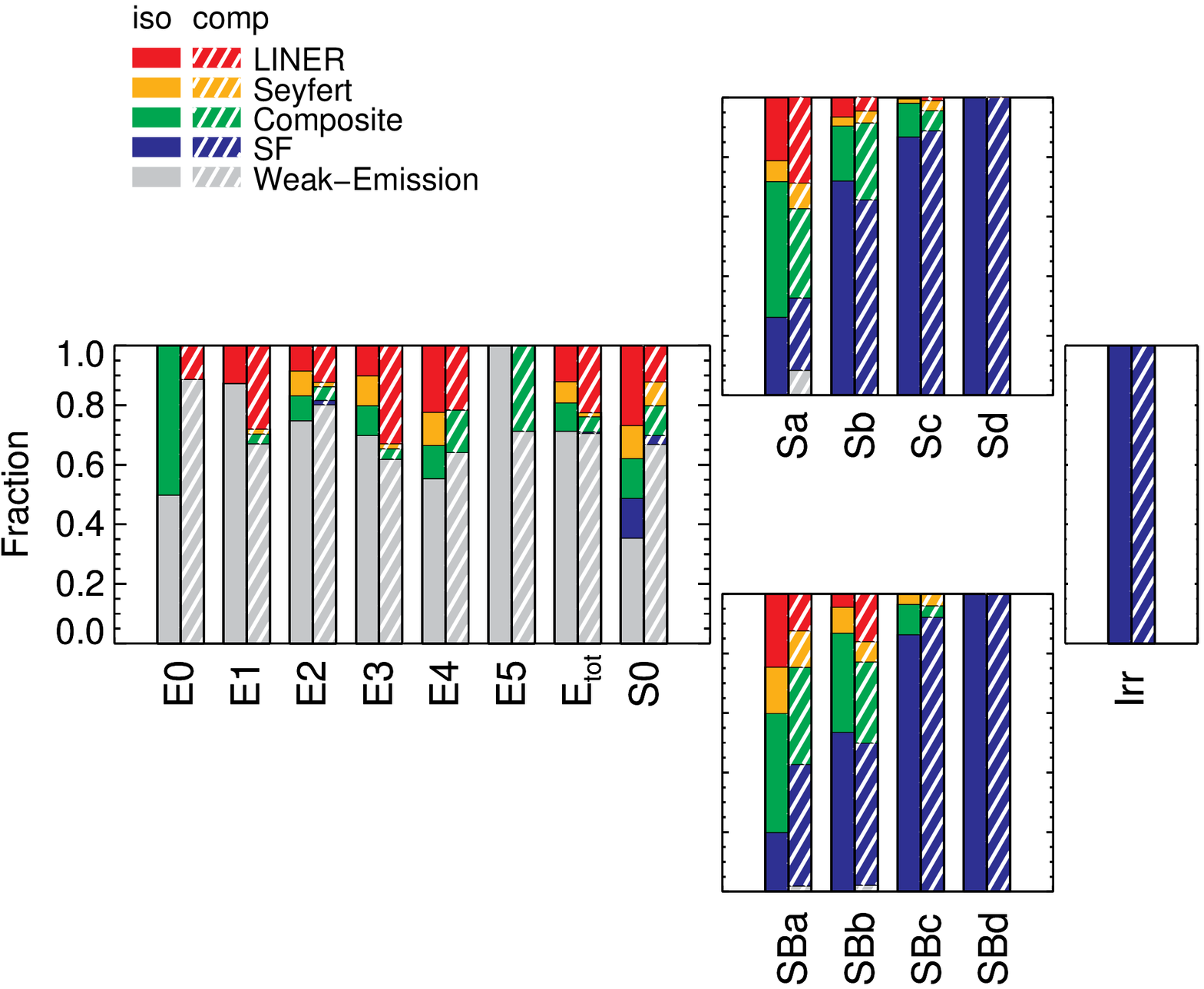}
\caption{Fraction of galaxies according to their emission line diagnostic. For each morphological type, the right bar is the comparison sample and the left bar is the isolated galaxies.}
\label{fig:bpt_fraction}
\end{figure}

%--------------- BPT table
\begin{deluxetable*}{lccccccccccccc}
%\tabletypesize{\scriptsize}
\tablecaption{Emission-line classification results}
\tablewidth{0pt}
\tablehead{
\colhead{Classification} &
\colhead{N} &
\colhead{$E_{tot}$} &
\colhead{S0} &
\colhead{Sa} &
\colhead{Sb} &
\colhead{Sc} &
\colhead{Sd} &
\colhead{SBa} &
\colhead{SBb} &
\colhead{SBc} &
\colhead{SBd} &
\colhead{Irr} &
\colhead{Total}} \\

\startdata
\multirow{2}*[-0.5ex]{Emission-line galaxies\tablenotemark{a}}
 & 911 & 0.75 & 1.81 & 3.56 & 18.06 & 20.12 & 0.75 & 2.81 & 4.25 & 3.25 & 0.12 & 1.19 & 56.69 \\[1pt]
 & (796) & (3.20) & (1.67) & (3.25) & (13.70) & (11.42) & (0.36) & (2.44) & (2.18) & (1.27) & (0.20) & (0.61) & (40.28) \\
\\

\multirow{2}*[-0.5ex]{-- Star-forming}
 & 633 & 0.00 & 0.38 & 0.94 & 12.94 & 17.44 & 0.75 & 0.56 & 2.25 & 2.75 & 0.12 & 1.19 & 39.31 \\[1pt]
 & (486) & (0.05) & (0.15) & (0.86) & (8.93) & (10.15) & (0.36) & (1.01) & (1.07) & (1.17) & (0.20) & (0.61) & (24.56) \\
\\

\multirow{2}*[-0.5ex]{-- Composite}
 & 175 & 0.25 & 0.38 & 1.62 & 3.38 & 2.38 & 0.00 & 1.12 & 1.44 & 0.38 & 0.00 & 0.00 & 10.94 \\[1pt]
 & (157) & (0.56) & (0.51) & (1.07) & (3.60) & (0.76) & (0.00) & (0.81) & (0.61) & (0.05) & (0.00) & (0.00) & (7.97) \\
\\

\multirow{2}*[-0.5ex]{-- Seyfert}
 & 40 & 0.19 & 0.31 & 0.25 & 0.56 & 0.25 & 0.00 & 0.44 & 0.38 & 0.12 & 0.00 & 0.00 & 2.50 \\[1pt]
 & (46) & (0.15) & (0.41) & (0.30) & (0.56) & (0.41) & (0.00) & (0.30) & (0.15) & (0.05) & (0.00) & (0.00) & (2.33) \\
\\

\multirow{2}*[-0.5ex]{-- LINER}
 & 63 & 0.31 & 0.75 & 0.75 & 1.19 & 0.06 & 0.00 & 0.69 & 0.19 & 0.00 & 0.00 & 0.00 & 3.94 \\[1pt]
 & (107) & (2.44) & (0.61) & (1.01) & (0.61) & (0.10) & (0.00) & (0.30) & (0.36) & (0.00) & (0.00) & (0.00) & (5.43) \\
\\

\multirow{2}*[-0.5ex]{Weak-emission\tablenotemark{b}}
 & 46 & 1.88 & 1.00 & 0.00 & 0.00 & 0.00 & 0.00 & 0.00 & 0.00 & 0.00 & 0.00 & 0.00 & 2.88 \\[1pt]
 & (229) & (7.76) & (3.40) & (0.30) & (0.05) & (0.00) & (0.00) & (0.05) & (0.05) & (0.00) & (0.00) & (0.00) & (11.62) \\
\\

\multirow{2}*[-0.5ex]{Unclear\tablenotemark{c}}
 & 648 & 7.56 & 8.75 & 2.44 & 8.38 & 9.25 & 0.06 & 1.63 & 1.63 & 0.75 & 0.00 & 0.00 & 40.43 \\[1pt]
 & (948) & (17.96) & (13.25) & (4.42) & (5.88) & (3.19) & (0.00) & (2.08) & (1.07) & (0.25) & (0.00) & (0.00) & (48.10) \\
\\

\multirow{2}*[-0.5ex]{Total\tablenotemark{d}}
 & 1605 & 10.19 & 11.56 & 6.00 & 26.44 & 29.37 & 0.81 & 4.44 & 5.88 & 4.00 & 0.12 & 1.19 & 100.00 \\[1pt]
 & (1973) & (28.92) & (18.32) & (7.97) & (19.63) & (14.61) & (0.36) & (4.57) & (3.30) & (1.52) & (0.20) & (0.61) & (100.00)

\enddata
\tablecomments{The upper and lower rows for each morphology class indicate the percent of isolated and comparison sample galaxies satisfying $10<\sigma_{eff}<400$ \kms, respectively.}
\tablenotetext{a}{$A/N \geq 3$ for \NII, H$\alpha$, \OIII and H$\beta$ emission lines. Emission-line galaxies consist of galaxies in the SF, Composite, Seyfert and LINER region.}
\tablenotetext{b}{$A/N \leq 1$.}
\tablenotetext{c}{Rest galaxies that are not included in emission-line galaxies nor weak-emission galaxies.}
\tablenotetext{d}{Summation of emission-line, weak-emission and unclear galaxies.}

\label{tab:bpt}
\end{deluxetable*}

\subsection{BPT Diagram}
\citet[][herafter BPT]{bpt81} demonstrated that several emission-line flux ratios are useful to classify emission line galaxies depending on their excitation mechanism. This diagnostic method has been revised by several studies \citep{vei87,kew01,kau03,kew06}. We used the demarcation lines suggested by \citet{kew01}, \citet{kau03} and \citet{sch07a} in order to categorize strong emission line galaxies into star forming, composite, Low Ionization Nuclear Emission-line Region (LINER) and Seyfert types. Figure~\ref{fig:bpt} shows the BPT diagram of isolated galaxies (upper panels) and comparison sample galaxies (lower panels) with $A/N \ge 3$ for all four emission lines ([\mbox{O {\sc iii}}] $\lambda$5007, H$\beta$, [\mbox{N {\sc ii}}] $\lambda$6584, and H$\alpha$). Here, $A/N$ is a ratio of the best fitting amplitude of emission line to residual noise of $\gandalf$ which is a similar to $S/N$. In this figure, the size and color of each symbol corresponds to the supermassive black hole mass and the (g--r)$_0$ color of the host galaxy, respectively. Black hole mass is calculated using the M$_{\rm BH}$--$\sigma$ relation in \citet{gul09} for galaxies with $10<\sigma_{eff}<400$ \kms.

Most of the galaxies with strong emission lines are late-type galaxies as shown in Figure~\ref{fig:bpt}. The contours of late-type galaxies (right panels) are predominantly found in the star forming region, and slightly extended to the composite region. There is no significant difference between isolated and comparison galaxies. Furthermore, the trends that the black hole mass is larger in the AGN region while color is bluer in the star forming region are same for isolated and comparison sample galaxies.

The distributions of early-type galaxies in the isolated and comparison samples on the BPT diagram seem different (left-hand panels). For isolated galaxies, emission-line early-type galaxies are more uniformly dispersed. It is notable that a large fraction of emission-line early-type galaxies in the star forming and composite region are lenticular galaxies (see the early-type galaxies in Figure~\ref{fig:bpt_fraction} and Table~\ref{tab:bpt}). Meanwhile, many of the emission-line early-type galaxies in the comparison sample are classified as LINER, which is consistent with previous studies \citep{hec80,kew06,sch07a}. LINERs are generally considered to be AGNs \citep{hec80,ho93}, indicating a higher fraction of AGNs in the early-type galaxies of the comparison sample. But we should bear in mind that there have been many debates on the excitation mechanism of LINER features. It may also be caused by the star formation \citep{ter85,shi92}, or old Post-AGB stars \citep{sar10,sin13}. \citet{sar10} in particular demonstrated that because the AGN activity is confined to the galactic center, other mechanisms such as star formation may affect the classification of ``true'' LINER in the SDSS, whose aperture size is larger than AGN sources.

The fraction of the galaxies in each region is presented in Figure~\ref{fig:bpt_fraction}. LINER, Seyfert, composite and star forming region are the same as shown in Figure~\ref{fig:bpt} and weak-emission galaxies are when $A/N \le 1$ for all 4 emission lines. Galaxies which are not weak-emission galaxies neither emission-line galaxies are classified as unclear-type galaxies and excluded in Figure~\ref{fig:bpt_fraction}. A considerable number of elliptical galaxies are weak-emission galaxies, and there is almost no weak-emission galaxies in late-type galaxies, as we would expect for star forming galaxies. The different subclasses of elliptical galaxies do not show any significant trend, although the LINER fractions in the comparison sample are slightly higher for all subclasses. For late-type galaxies, the fraction of star forming galaxies increases from S(B)a to S(B)d, as they become increasingly late-type. Meanwhile the fraction in AGN and composite region decreases along the Hubble sequence for isolated and comparison sample galaxies. The detailed information is summarized in Table~\ref{tab:bpt}.

% --------------------------------------------------------------------------------

% Figure - H alpha
\begin{figure}
\centering
\includegraphics[width=0.5\textwidth]{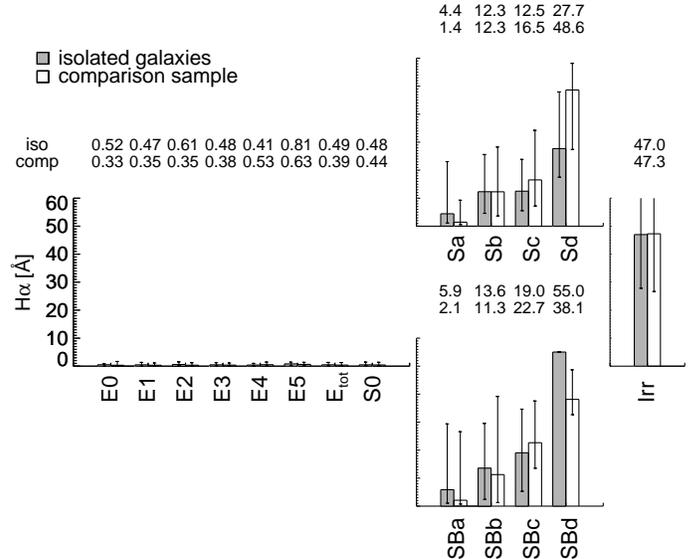}
\caption{The H$\alpha$ equivalent width of non-AGN galaxies. The format is the same as that of Figure~\ref{fig:cr}.}
\label{fig:h_alpha}
\end{figure}

% Figure -SAM model
\begin{figure*}
\begin{center}
\includegraphics[width=0.42\textwidth, angle = 270]{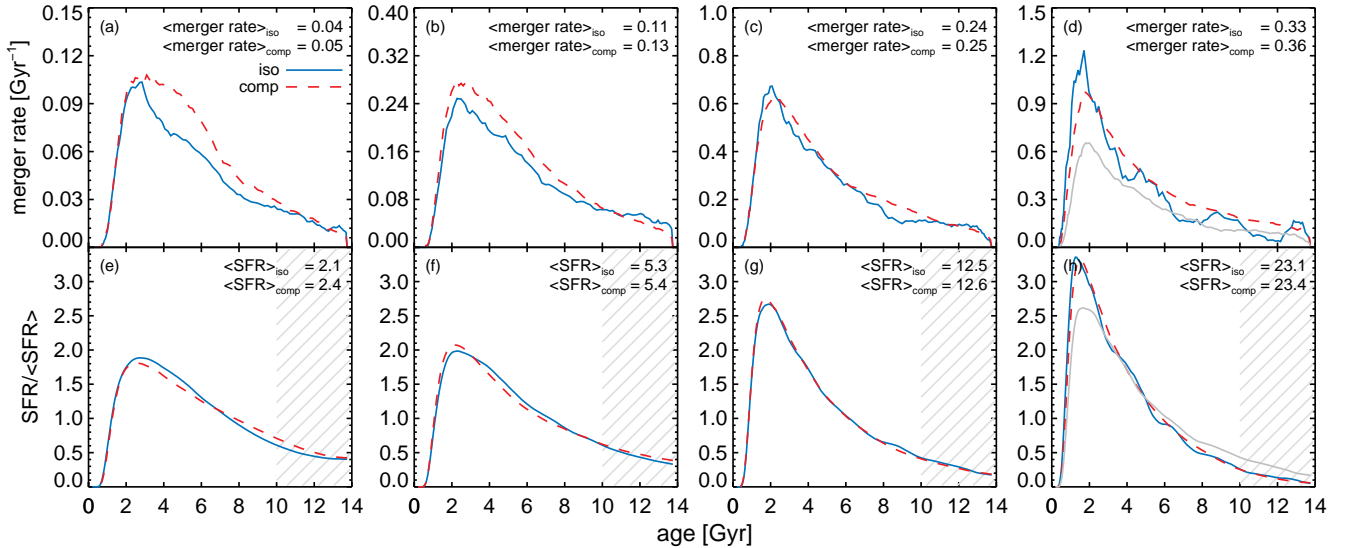}
\caption{Merger rate and normalized SFR. Abscissa: the age of universe; Ordinate: the merger rate (upper panels) and normalized SFR (lower panels). From left to right, the median value of stellar mass changes from $10^{10.2}, 10^{10.5}, 10^{10.8}$ to 10$^{11.1}\ M_\odot$, which corresponds to the typical masses of Sc, Sb, Sa-type spiral and elliptical galaxies in our  observational data. Blue solid lines and red dashed lines correspond to isolated and comparison galaxies, respectively. The mean values of the merger rate and SFR are shown at the top right corner of each panel. 
The gray solid lines in the right-most panels indicate the profiles for the isolated galaxies that are 50\% less massive than the galaxies in these panels, representing the models that have similar mass to the isolated elliptical galaxies (10$^{10.9}\ M_\odot$) in the SDSS data. See text for more information.
The hatched regions in the bottom panels roughly indicate the age range (past 4\,Gyr) that strongly affects the luminosity-weighted age based on H$\beta$ index.}
\label{fig:history}
\end{center}
\end{figure*}

\subsection{Star formation activities}
% H-alpha emission
Young stars ionize gas in starforming regions, forming \mbox{H {\sc ii}} regions. H$\alpha$ emission is produced in these regions, and this has been studied frequently in the past \citep[e.g.,][]{coh76,ken83,gav91,ken92,ryd94,gal95}. Figure~\ref{fig:h_alpha} shows the H$\alpha$ emission line strength in equivalent width for non-AGN host galaxies which include star forming, composite, unclear-type and weak emission galaxies. The median EW values of the H$\alpha$ emission line are very small in early-type galaxies and there is no significant difference between isolated galaxies and comparison sample, because star forming early-type galaxies only account for a small fraction.

In late-type galaxies, the strength of the H$\alpha$ emission line increases from early to late-type spirals for both isolated and comparison galaxies. \citet{bus87} and \citet{ken87} report that the H$\alpha$ emission of interacting galaxies is stronger than that of an isolated galaxy. In addition, \citet{kav14} claims that minor mergers can enhance the star formation in disk galaxies, and this enhancement is more significant in late-type spirals than in early-type spiral galaxies. In Figure~\ref{fig:h_alpha}, we show that the median values of H$\alpha$ emission line strength of isolated galaxies are not significantly different from those of comparison galaxies for S(B)a and S(B)b galaxies. S(B)c and Sd galaxies (except for SBd galaxies due to the small number) in the comparison sample may have slightly larger values than isolated ones, but the difference is not as conspicuous as reported in \citet{kav14}, perhaps partly due to our large error bars.

% --------------------------------------------------------------------------------

\section{Discussion}
\label{sec:discussion}
\subsection{Comparison with Semi-analytic Models}

The most notable differences between the isolated and comparison samples can be summarized as follows: isolated galaxies are much more likely to be late type in morphology, and isolated early-type galaxies are 50\% less massive and 20\% younger than their counterparts. The most obvious candidate for the origin of such differences may be found in the merger history of galaxies.

Semi-analytic models can provide a comprehensive picture of galaxy evolution history according to different density environments  \citep[e.g.,][]{wan13,jun14}. In order to investigate the origin of the various galaxy properties in different environments, we use a semi-analytic model developed by \citet{lee13} based on a cosmological N-body volume simulation of structure formation run using GADGET-2 \citep{spr05}. For the volume, we adopt the cosmological parameters derived from the {\em Wilkinson Microwave Anisotropy Probe seven year observations} \citep{kom11}, $\Omega_{\rm m}$=0.272, $\Omega_{\Lambda}=0.728$, and $h=0.704$. The periodic cube size of the volume is $200 h^{-1}$Mpc on a side with $1024^3$ collision-less particles. We generated a halo catalog by identifying substructures from friends-of-friends (FOF) groups using SUBFIND \citep{spr01}. Then, halo merger trees were constructed from the halo catalog by a tree building algorithm described in \citet{jun14}. We used the merger trees as an input of the semi-analytic model.  Further details of prescriptions for baryonic physics can be found in \citet{lee13}.

The results of the semi-analytic model are shown in Figure~\ref{fig:history}. From left to right panels, the stellar mass is varied following $10^{10.2}$ (panels a and e), $10^{10.5}$ (b and f), $10^{10.8}$ (c and g), and $10^{11.1} M_\odot$ (d and h) and these stellar masses roughly correspond to the median values of the Sc, Sb, Sa-type spiral and elliptical galaxies in our sample (Figure ~\ref{fig:mass}). Also, the median bulge-to-total ratio of the galaxies in each panel is roughly 0.1, 0.3, 0.5 and 0.8, respectively, which is roughly consistent with the observational data \citep[e.g.,][]{hud10}. The galaxies used in the semi-analytic model satisfy the density criteria of our sampling method considering the same stellar mass cut ($M_\ast \gtrsim 10^9\ M_\odot$) as the one used in our volume limitation on the SDSS data.

Upper panels of Figure~\ref{fig:history} are the averaged evolution of the merger rate for isolated (blue solid line) and comparison galaxies (red dashed line) over the age of the Universe. Only mergers in a ``direct progenitor'' are taken into account when we calculate the merger rate\footnote{The overall trend remains the same even when we use ``all progenitors'' instead of ``direct progenitors" in the analysis.}. Here, a direct progenitor means the galaxy in the most massive halo in a merger tree at a given epoch\citep{lee13}. Also, only mergers with a stellar mass ratio greater than 1:10 (which may significantly affect the galaxy's evolution) are considered to derive the merger rate. In each panel, the mean merger rate is indicated in units of Gyr$^{-1}$ per galaxy and the mean merger rates increase as galaxies become more massive, as expected. Interestingly, however, the mean merger rates (the values in the top of each panel) are similar within each panel (that is, effectively for a given stellar mass) regardless of the density of the environment. The merger-rate profiles of the two samples also seem similar to each other, perhaps with the exception of the first panel (lowest mass sample).

Spiral galaxies do not show much dependence on environment. But this is sensible from the observational point of view. Spirals, whether in isolation or in dense regions, cannot have had a significant number of recent major mergers, because otherwise they would not have remained as spirals to date. This is supported by our model as well. Note that the merger rates for later type spirals (e.g., panel a) are very low: typically only one important (1:10 or larger) merger is expected for the last 10 billion years for these galaxies. 

As mentioned above, earlier type spiral galaxies have gradually higher merger rates (top rows), and hence, if we consider all spiral galaxies together from Sa to Sd-type, the combined merger-rate profile of isolated galaxies will be lower than that of comparisons due to the high fraction of later-type spiral galaxies in isolated systems.

The similar merger rates of elliptical galaxies of the two samples (panel d) seem contrary to the expectation that the elliptical galaxies in denser regions may suffer significantly more mergers than isolated elliptical galaxies. However, the observed data suggest that isolated elliptical galaxies are 50\% less massive than their counterparts in the comparison sample, and this must be considered in the comparison. We present the profiles for such less massive elliptical galaxies from the models in panel (d) as gray solid lines. The less massive elliptical galaxies had lower merger rates than more massive elliptical galaxies (red and blue lines). So, in this panel we should consider the red dashed line a representation for the elliptical galaxies in our comparison sample and the gray solid line for the isolated ellipticals.

SFRs normalized by the mean SFRs are shown in the lower panels. The SFR presented here is the cumulative SFR of all progenitors (galaxies in the host halo and subhalos). The shape of the SFR evolution seems to follow that of the merger rate, which implies that mergers between galaxies play an important role in the star formation history in the framework of semi-analytic models. Like the merger rate, the mean SFRs in units of M$_\odot$/yr and the profiles of normalized SFRs are similar for isolated and comparison galaxies in the same mass range. 

The luminosity-weighted age of a galaxy is sensitive to the star formation occurring during the past few Gyr. Therefore, the similar normalized SFRs of isolated and comparison galaxies during the past 3--4 Gyr (hatched region) can lead to the similar luminosity-weighted ages for a fixed stellar mass. However, as mentioned above, the comparison galaxies in the highest mass bin should be compared with less massive isolated galaxies due to the difference in stellar mass. In panel (h), the gray solid line (isolated galaxies with $10^{10.9}\ M_\odot$) shows slightly higher normalized SFR in recent time than the red dashed line (comparison galaxies with $10^{11.1}\ M_\odot$). This means that there are larger amounts of accumulated star formation in isolated galaxies during the last few Gyr and therefore, the isolated elliptical galaxies would be measured to be younger than the comparison sample in luminosity-weighted age. The results are compatible with those from the absorption-line analysis -- that isolated elliptical galaxies appear younger than their counterparts while spiral galaxies in the isolated and comparison samples have similar luminosity-weighted ages for a given morphology. In this interpretation, the tendency in the luminosity-weighted ages of galaxies, at least in the local Universe, may be viewed as a galaxy mass effect rather than environmental effect \citep{pen10,tho10,jun14}.

% --------------------------------------------------------------------------------

% Figure -Bias test
\begin{figure}
\begin{center}
\includegraphics[width=0.43\textwidth]{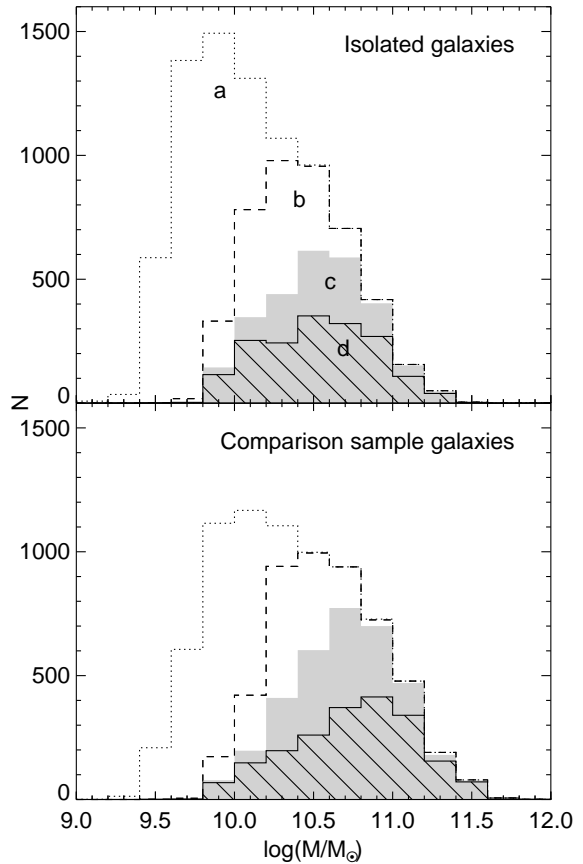}
\caption{Stellar mass distribution for isolated (upper) and comparison sample galaxies (lower). a ($M_{\rm r}\leq-18.68$), b ($M_{\rm r}\leq-19.68$), c (applying the cut in \citet{oh13}) and d (excluding the unknown type) histograms are for different samples. See the text for details.}
\label{fig:bias}
\end{center}
\end{figure}

\subsection{Sampling Bias}
In this section, we discuss possible bias that could affect our works. The situations where bias could be introduced are the following:

\begin{description}
\item[Case 1] The selection of the galaxies with $M_{\rm r}\leq-19.68$ from the volume-limited sample ($M_{\rm r}\leq-18.68$).
\item[Case 2] The use of the criteria in \citet{oh13}.
\item[Case 3] The classification of morphology into the unknown class during the visual inspection.
\end{description}

Figure~\ref{fig:bias} shows the distribution of the stellar mass for isolated (upper panel) and comparison sample galaxies (lower panel). Different histogram styles indicate different samples. The dotted-line open histogram (a) is for the sample with $M_{\rm r}\leq-18.68$. The dashed-line open histogram (b) is for the sample with $M_{\rm r}\leq-19.68$. The filled histogram (c) is for the sample after applying the cut in \citet{oh13}. The hatched histogram (d) is for the sample after excluding the unknown class through the visual inspection. The overall distributions of the comparison sample are slightly shifted toward high mass compared to those of isolated galaxies. Some low-mass isolated (or comparison) galaxies in the dotted-line open histogram (a) would be misclassified because of the inhomogeneous sampling (see Section~\ref{sec:sample}). Case 1 causes sample a to change to sample b, Case 2 is from sample b to c, and Case 3 is from sample c to d.

Since Sd-type spiral galaxies have the lowest mass among spiral galaxies, they are likely to be abundant at $<10^{10}$ $M_{\odot}$. In Case 1, most galaxies with a mass range of $\sim$$10^{9}$--$10^{10}$ $M_{\odot}$ are missed, so many Sd-type galaxies would be removed, only leaving a small number of massive Sd-type galaxies. In addition, among Sc-type galaxies, the galaxies with low mass would also be removed due to its median stellar mass being close to $10^{10}$ $M_{\odot}$. This would make Sc-type spiral galaxies biased toward massive ones. In Case 2, the small and apparently flat galaxies are likely to be excluded. Thus, there is a chance that Sb-type and Sc-type spirals, as well as E6 and E7 elliptical galaxies, are missed. As discussed in Section~\ref{sec:morph}, S0 galaxies are not easy to classify and thus, S0 galaxies and complex Sa-type and elliptical galaxies would account for a large fraction of the unknown class in our study (Section~\ref{sec:morph_compare}). As a result, some galaxies with the mass range of $\sim$$10^{10}$--$10^{11}$ $M_{\odot}$ can be excluded in Case 3, and some late-type spiral galaxies are missed. Hence, through Case 1, 2 and 3, spiral galaxies seem to be affected not only in number fraction but in stellar mass, whereas elliptical galaxies with mass $\gtrsim10^{11}$ $M_{\odot}$ are less affected. To summarize, although we have found that spiral galaxies are rich in isolated galaxies (Figure~\ref{fig:fraction}), the fraction of spiral galaxies would be underestimated due to the sampling bias. The stellar mass in spiral galaxies could be somewhat biased toward massive galaxies in Figure~\ref{fig:mass}, but this occurs to a lesser degree in elliptical galaxies.

% --------------------------------------------------------------------------------

\section{Summary}
\label{sec:summary}
We have used 1,644 isolated galaxies and 1,994 comparison galaxies, drawn from the SDSS DR7, in the redshift range of $0.025 < z < 0.044$, to investigate the properties of isolated galaxies along the Hubble morphological types. Our study focuses on not only photometric  properties, but also spectroscopic characteristics of isolated galaxies, with the help of the OSSY catalog which provides accurate line measurements. The main results of this study can be summarized as follows.

\begin{enumerate}[leftmargin=*] \itemsep.5pt
\item To study isolated galaxies, we construct the isolated galaxy sample and its comparison group based on the number of companion galaxies. We define the galaxies without any companion galaxy within a 2 Mpc sphere as isolated galaxies and those with at least two companions within 1 Mpc as comparison sample galaxies. By choosing galaxies with $M_{\rm r}\leq-19.68$ from our volume-limited sample, we ensure we can detect companions down to a mass ratio of $\sim$1:3 for our faintest galaxies. This choice provides us with comparable numbers of galaxies between the isolated and comparison sample. 

\item Isolated galaxies are distributed widely and even located in high-density (by long-range density measurement) regions, whereas comparison galaxies are mainly positioned in high-density areas. We have also found that there are a small number of void galaxies in our isolated galaxy sample as well as in the comparison set. This implies that the isolated galaxies in our study are primarily influenced by the local density environment, rather than the global one.

\item Morphologies of the two samples are classified through a visual inspection and the fractions along the Hubble sequence of isolated and comparison galaxies show different distributions. Only 21.2\% of isolated galaxies are early type whereas 46.7\% are early type in the comparison sample. Naturally, spiral galaxies are much more common in isolated systems (77.6\%) than in comparison ones (52.5\%). This result is consistent with the morphology-density relation \citep{dre80}. Subclasses (E0--E5) of elliptical galaxies in the two samples have similar distributions, because we do not consider the intrinsic shapes. The fractions of barred spiral galaxies seem similar regardless of the density of the environment.

\item Elliptical galaxies of both isolated and comparison galaxies form the red sequence on the color-magnitude diagram, but, on the whole, isolated ones ($M_{\rm r} = -21.36$) appear fainter than their comparisons ($M_{\rm r} = -21.73$). Unbarred spiral galaxies of both samples are in the blue cloud and overall distributions are similar. Barred spiral galaxies show bimodal distributions without any distinct difference between the isolated and comparison samples.

\item For elliptical galaxies, there is a discrepancy in stellar mass between the isolated and comparison sample galaxies. The median of the stellar mass in isolated elliptical galaxies ($10^{10.92}\ M_\odot$) is lower than that of their comparisons ($10^{11.10}\ M_\odot$) by 50\%. On the other hand, spiral galaxies do not show any significant difference in their stellar mass between the two samples {\em within subclasses}. The same trends are observed in the distributions of their velocity dispersion.

\item The median absorption line strengths including H$\beta$, Fe5270 and Mgb between isolated galaxies and their comparisons are similar for all morphological types.

\item We used the TMJ stellar population model \citep{tho11} which can help to break the degeneracy between the age and metallicity of a galaxy. Elliptical galaxies show a discrepancy between isolated and comparison galaxies in the median luminosity-weighted age by 20\% and there is no obvious distinction in the median luminosity-weighted metallicity or [$\alpha$/Fe]. Spiral galaxies in the two samples do not show any significant difference in properties of stellar populations.

\item We have not found any significant difference in emission-line properties between isolated and comparison samples that can be explained in a straightforward way.

\end{enumerate}

We constructed a sample of isolated galaxies following conventional approaches based on local density. It is somewhat surprising that isolated galaxies do not show much difference in observable properties, but this is largely supported by the current hierarchical prediction. It seems that the ``isolated'' galaxies defined this way are not an extreme population of galaxies after all, as long as they are part of the same large scale structures.

We hereby provide the public with the tables that contain the list of all of our sample galaxies and their properties measured from the observed data and derived from the comparison with stellar population models.

% --------------------------------------------------------------------------------

\acknowledgments

SKY acknowledges support from the National Research Foundation of Korea (Doyak 2014003730). Numerical simulations were performed using the KISTI supercomputer under the programme of KSC-2013-C3-015. This study was performed under the DRC collaboration between Yonsei University and the Korea Astronomy and Space Science Institute.

%----------Summary Table 1
\begin{deluxetable*}{lcccccccc}
\tablecaption{Basic properties of the Early-type galaxies}
\tablewidth{0pt}
\tablehead{
\colhead{} &
\colhead{E0} &
\colhead{E1} &
\colhead{E2} &
\colhead{E3} &
\colhead{E4} &
\colhead{E5} &
\colhead{$E_{tot}$} &
\colhead{S0}}

\startdata
\multirow{2}*[-0.5ex]{\%\tablenotemark{a}}
 & 0.36 & 2.25 & 2.55 & 2.86 & 1.52 & 0.36 &  9.91 & 11.25 \\[1pt]
 & (1.25) & (7.57) & (8.83) & (6.97) & (2.91) & (1.05) & (28.59) & (18.10) \\
\\

\multirow{2}*[-0.5ex]{Cr\tablenotemark{b}}
& 3.21$^{+0.04}_{-0.08}$ & 3.14$^{+0.23}_{-0.13}$ & 3.18$^{+0.16}_{-0.19}$ & 3.24$^{+0.16}_{-0.19}$ & 3.24$^{+0.21}_{-0.19}$ & 3.14$^{+0.06}_{-0.09}$ & 3.20$^{+0.18}_{-0.18}$ & 3.04$^{+0.23}_{-0.28}$ \\[4pt]
& (3.18$^{+0.19}_{-0.16}$) & (3.19$^{+0.19}_{-0.17}$) & (3.25$^{+0.14}_{-0.24}$) & (3.23$^{+0.16}_{-0.24}$) & (3.17$^{+0.18}_{-0.25}$) & (3.22$^{+0.12}_{-0.23}$) & (3.22$^{+0.16}_{-0.22}$) & (3.07$^{+0.27}_{-0.29}$) \\
\\

\multirow{2}*[-0.5ex]{\fracDeV\tablenotemark{b}}
& 0.98$^{+0.02}_{-0.02}$ & 1.00$^{+0.00}_{-0.06}$ & 1.00$^{+0.00}_{-0.09}$ & 1.00$^{+0.00}_{-0.04}$ & 1.00$^{+0.00}_{-0.04}$ & 0.98$^{+0.02}_{-0.08}$ & 1.00$^{+0.00}_{-0.07}$ & 1.00$^{+0.00}_{-0.13}$ \\[4pt]
& (1.00$^{+0.00}_{-0.06}$) & (1.00$^{+0.00}_{-0.07}$) & (1.00$^{+0.00}_{-0.05}$) & (1.00$^{+0.00}_{-0.07}$) & (1.00$^{+0.00}_{-0.14}$) & (1.00$^{+0.00}_{-0.13}$) & (1.00$^{+0.00}_{-0.07}$) & (1.00$^{+0.00}_{-0.13}$) \\
\\

\multirow{2}*[-0.5ex]{$(g-r)_0$\tablenotemark{d}}
& 0.77$^{+0.01}_{-0.07}$ & 0.76$^{+0.03}_{-0.02}$ & 0.75$^{+0.03}_{-0.04}$ & 0.76$^{+0.03}_{-0.04}$ & 0.75$^{+0.03}_{-0.02}$ & 0.79$^{+0.02}_{-0.06}$ & 0.76$^{+0.03}_{-0.04}$ & 0.75$^{+0.03}_{-0.04}$ \\[4pt]
& (0.77$^{+0.03}_{-0.02}$) & (0.77$^{+0.02}_{-0.03}$) & (0.77$^{+0.03}_{-0.03}$) & (0.77$^{+0.03}_{-0.03}$) & (0.77$^{+0.03}_{-0.03}$) & (0.75$^{+0.02}_{-0.03}$) & (0.77$^{+0.03}_{-0.03}$) & (0.76$^{+0.03}_{-0.03}$) \\
\\

\multirow{2}*[-0.5ex]{$M_{\rm r}$\tablenotemark{e}}
& $-21.33^{+0.56}_{-0.39}$ & $-21.42^{+0.62}_{-0.71}$ & $-21.25^{+0.58}_{-0.51}$ & $-21.39^{+0.54}_{-0.47}$ & $-21.48^{+0.65}_{-0.40}$ & $-21.47^{+0.47}_{-0.42}$ & $-21.36^{+0.65}_{-0.53}$ & $-20.92^{+0.43}_{-0.42}$ \\[4pt]
& ($-21.79^{+0.32}_{-0.38}$) & ($-21.59^{+0.56}_{-0.60}$) & ($-21.80^{+0.55}_{-0.62}$) & ($-21.78^{+0.46}_{-0.66}$) & ($-21.73^{+0.56}_{-0.50}$) & ($-21.66^{+0.37}_{-0.47}$) & ($-21.73^{+0.53}_{-0.59}$) & ($-21.05^{+0.45}_{-0.43}$) \\
\\

\multirow{2}*[-0.5ex]{log($M_{\ast}$/$M_{\odot}$)\tablenotemark{f}}
& 10.90$^{+0.26}_{-0.11}$ & 10.96$^{+0.27}_{-0.29}$ & 10.87$^{+0.30}_{-0.19}$ & 10.92$^{+0.22}_{-0.18}$ & 10.94$^{+0.34}_{-0.12}$ & 10.99$^{+0.26}_{-0.14}$ & 10.92$^{+0.26}_{-0.23}$ & 10.76$^{+0.18}_{-0.18}$ \\[4pt]
& (11.12$^{+0.22}_{-0.18}$) & (11.04$^{+0.22}_{-0.26}$) & (11.14$^{+0.23}_{-0.28}$) & (11.11$^{+0.20}_{-0.25}$) & (11.12$^{+0.23}_{-0.23}$) & (11.02$^{+0.18}_{-0.18}$) & (11.10$^{+0.23}_{-0.25}$) & (10.83$^{+0.18}_{-0.19}$)

\enddata
\tablecomments{The upper and lower rows for each morphology class indicate the values of properties of isolated and comparison sample galaxies, respectively}
\tablenotetext{a}{The percentage of galaxies for each morphology.}
\tablenotetext{b}{Concentration index, C$_r$=PetroR90/PetroR50.  Upper row indicates the median value and 1$\sigma$ dispersion of isolated galaxies. Values in lower row with parenthesis is for comparison galaxies.}
\tablenotetext{c}{\fracDeV, de Vaucouleurs fraction of the $r$-band from the SDSS photometric pipelines.}
\tablenotetext{d}{Extinction-corrected and $k$-corrected (g-r) color using the petrosian magnitude.}
\tablenotetext{e}{Extinction-corrected and $k$-corrected $r$-band absolute magnitude.}
\tablenotetext{f}{Logarithmic scale of the stellar mass \citep{bel03}.}

\label{tab:catalogone}
\end{deluxetable*}

%----------Summary Table 2
\setlength{\tabcolsep}{0.1cm}
\begin{deluxetable*}{lccccccccc}
\tablecaption{Basic properties of the Late-type and Irregular galaxies}
\tablewidth{0pt}
\tablehead{
\colhead{} &
\colhead{Sa} &
\colhead{Sb} &
\colhead{Sc} &
\colhead{Sd} &
\colhead{SBa} &
\colhead{SBb} &
\colhead{SBc} &
\colhead{SBd} &
\colhead{Irr}}

\startdata
\multirow{2}*[-0.5ex]{\%}
 & 5.84 & 26.28 & 30.11 & 0.91 & 4.32 & 5.78 & 4.32 & 0.12 & 1.16 \\[1pt]
 & (7.87) & (19.66) & (15.25) & (0.40) & (4.51) & (3.26) & (1.55) & (0.20) & (0.60) \\
\\

\multirow{2}*[-0.5ex]{Cr}
& 2.73$^{+0.35}_{-0.32}$ & 2.23$^{+0.31}_{-0.24}$ & 2.11$^{+0.26}_{-0.18}$ & 2.10$^{+0.31}_{-0.16}$ & 2.46$^{+0.42}_{-0.31}$ & 2.16$^{+0.30}_{-0.19}$ & 2.10$^{+0.22}_{-0.16}$ & 2.16$^{+0.00}_{-0.00}$ & 2.61$^{+0.40}_{-0.41}$ \\[4pt]
& (2.71$^{+0.39}_{-0.40}$) & (2.24$^{+0.31}_{-0.26}$) & (2.13$^{+0.26}_{-0.15}$) & (2.11$^{+0.28}_{-0.09}$) & (2.48$^{+0.39}_{-0.32}$) & (2.31$^{+0.30}_{-0.25}$) & (2.15$^{+0.26}_{-0.17}$) & (2.07$^{+0.05}_{-0.03}$) & (2.33$^{+0.31}_{-0.21}$) \\
\\

\multirow{2}*[-0.5ex]{\fracDeV}
& 0.84$^{+0.16}_{-0.36}$ & 0.18$^{+0.36}_{-0.18}$ & 0.12$^{+0.22}_{-0.12}$ & 0.09$^{+0.10}_{-0.06}$ & 0.83$^{+0.17}_{-0.49}$ & 0.46$^{+0.33}_{-0.32}$ & 0.26$^{+0.41}_{-0.16}$ & 0.28$^{+0.00}_{-0.00}$ & 0.35$^{+0.20}_{-0.34}$ \\[4pt]
& (0.82$^{+0.18}_{-0.31}$) & (0.24$^{+0.40}_{-0.21}$) & (0.15$^{+0.24}_{-0.15}$) & (0.02$^{+0.31}_{-0.02}$) & (0.89$^{+0.11}_{-0.18}$) & (0.69$^{+0.31}_{-0.46}$) & (0.36$^{+0.33}_{-0.25}$) & (0.25$^{+0.07}_{-0.04}$) & (0.26$^{+0.14}_{-0.26}$) \\
\\

\multirow{2}*[-0.5ex]{$(g-r)_0$}
& 0.67$^{+0.06}_{-0.05}$ & 0.56$^{+0.08}_{-0.10}$ & 0.49$^{+0.07}_{-0.09}$ & 0.39$^{+0.08}_{-0.05}$ & 0.67$^{+0.05}_{-0.06}$ & 0.60$^{+0.06}_{-0.10}$ & 0.49$^{+0.09}_{-0.12}$ & 0.37$^{+0.00}_{-0.00}$ & 0.37$^{+0.09}_{-0.06}$ \\[4pt]
& (0.70$^{+0.06}_{-0.06}$) & (0.57$^{+0.09}_{-0.09}$) & (0.48$^{+0.09}_{-0.08}$) & (0.33$^{+0.05}_{-0.05}$) & (0.70$^{+0.05}_{-0.06}$) & (0.61$^{+0.08}_{-0.08}$) & (0.47$^{+0.11}_{-0.06}$) & (0.36$^{+0.02}_{-0.03}$) & (0.37$^{+0.08}_{-0.02}$) \\
\\

\multirow{2}*[-0.5ex]{$M_{\rm r}$}
& $-21.36^{+0.46}_{-0.51}$ & $-20.81^{+0.50}_{-0.55}$ & $-20.59^{+0.58}_{-0.50}$ & $-20.16^{+0.25}_{-0.22}$ & $-21.29^{+0.46}_{-0.50}$ & $-21.05^{+0.52}_{-0.66}$ & $-20.43^{+1.10}_{-0.43}$ & $-20.29^{+0.00}_{-0.00}$ & $-20.25^{+0.52}_{-0.22}$ \\[4pt]
& ($-21.41^{+0.55}_{-0.61}$) & ($-20.90^{+0.67}_{-0.62}$) & ($-20.64^{+0.69}_{-0.47}$) & ($-20.12^{+0.24}_{-0.30}$) & ($-21.37^{+0.52}_{-0.66}$) & ($-21.26^{+0.54}_{-0.71}$) & ($-20.18^{+0.64}_{-0.19}$) & ($-20.44^{+0.06}_{-0.08}$) & ($-20.24^{+0.07}_{-0.06}$) \\
\\

\multirow{2}*[-0.5ex]{log($M_{\ast}$/$M_{\odot}$)}
& 10.84$^{+0.22}_{-0.24}$ & 10.48$^{+0.29}_{-0.29}$ & 10.34$^{+0.26}_{-0.27}$ & 10.08$^{+0.10}_{-0.08}$ & 10.80$^{+0.23}_{-0.24}$ & 10.64$^{+0.25}_{-0.36}$ & 10.28$^{+0.45}_{-0.28}$ & 10.09$^{+0.00}_{-0.00}$ & 10.07$^{+0.29}_{-0.15}$ \\[4pt]
& (10.89$^{+0.26}_{-0.29}$) & (10.55$^{+0.32}_{-0.30}$) & (10.35$^{+0.32}_{-0.25}$) & ( 9.96$^{+0.10}_{-0.09}$) & (10.88$^{+0.23}_{-0.29}$) & (10.74$^{+0.25}_{-0.39}$) & (10.15$^{+0.35}_{-0.13}$) & (10.14$^{+0.05}_{-0.07}$) & (10.04$^{+0.13}_{-0.05}$)

\enddata

\label{tab:catalogtwo}
\end{deluxetable*}
%----------Summary Table 3
\begin{deluxetable*}{lcccccccc}
\tablecaption{Basic properties of the Early-type galaxies}
\tablewidth{0pt}
\tablehead{
\colhead{} &
\colhead{E0} &
\colhead{E1} &
\colhead{E2} &
\colhead{E3} &
\colhead{E4} &
\colhead{E5} &
\colhead{$E_{tot}$} &
\colhead{S0}}

\startdata
\multirow{2}*[-0.5ex]{$\sigma_{eff}$\tablenotemark{a} (\kms)}
& 127.6$^{+64.5}_{-18.5}$ & 169.3$^{+23.0}_{-58.0}$ & 149.1$^{+52.3}_{-25.9}$ & 175.3$^{+30.4}_{-40.4}$ & 167.1$^{+51.3}_{-38.9}$ & 174.9$^{+23.3}_{-16.6}$ & 160.6$^{+44.7}_{-39.0}$ & 133.6$^{+36.0}_{-41.4}$ \\[4pt]
& (198.1$^{+35.0}_{-47.5}$) & (187.0$^{+40.4}_{-51.7}$) & (201.5$^{+34.2}_{-49.4}$) & (196.1$^{+36.4}_{-38.2}$) & (194.6$^{+35.5}_{-31.8}$) & (183.1$^{+47.1}_{-30.6}$) & (194.2$^{+38.2}_{-45.1}$) & (143.6$^{+45.1}_{-38.4}$) \\
\\

\multirow{2}*[-0.5ex]{H$\beta$\tablenotemark{b} (\AA)}
& 2.02$^{+0.15}_{-0.23}$ & 1.84$^{+0.16}_{-0.21}$ & 1.82$^{+0.30}_{-0.19}$ & 1.76$^{+0.29}_{-0.11}$ & 1.89$^{+0.41}_{-0.18}$ & 1.87$^{+0.02}_{-0.15}$ & 1.83$^{+0.28}_{-0.17}$ & 1.90$^{+0.33}_{-0.22}$ \\[4pt]
& (1.76$^{+0.17}_{-0.16}$) & (1.74$^{+0.21}_{-0.20}$) & (1.78$^{+0.17}_{-0.22}$) & (1.72$^{+0.20}_{-0.14}$) & (1.68$^{+0.24}_{-0.15}$) & (1.73$^{+0.19}_{-0.16}$) & (1.75$^{+0.20}_{-0.19}$) & (1.88$^{+0.30}_{-0.24}$) \\
\\

\multirow{2}*[-0.5ex]{Fe5270\tablenotemark{c} (\AA)}
& 2.79$^{+0.05}_{-0.09}$ & 2.84$^{+0.15}_{-0.21}$ & 2.72$^{+0.24}_{-0.29}$ & 2.81$^{+0.23}_{-0.32}$ & 2.76$^{+0.31}_{-0.27}$ & 2.92$^{+0.04}_{-0.49}$ & 2.80$^{+0.21}_{-0.31}$ & 2.80$^{+0.26}_{-0.27}$ \\[4pt]
& (2.93$^{+0.07}_{-0.26}$) & (2.83$^{+0.20}_{-0.20}$) & (2.87$^{+0.20}_{-0.24}$) & (2.84$^{+0.19}_{-0.24}$) & (2.84$^{+0.13}_{-0.19}$) & (2.82$^{+0.29}_{-0.25}$) & (2.85$^{+0.20}_{-0.23}$) & (2.86$^{+0.20}_{-0.24}$) \\
\\

\multirow{2}*[-0.5ex]{$<$Fe$>$\tablenotemark{d} (\AA)}
& 2.76$^{+0.07}_{-0.22}$ & 2.84$^{+0.22}_{-0.20}$ & 2.73$^{+0.23}_{-0.23}$ & 2.81$^{+0.21}_{-0.28}$ & 2.70$^{+0.37}_{-0.19}$ & 2.95$^{+0.05}_{-0.45}$ & 2.80$^{+0.23}_{-0.28}$ & 2.80$^{+0.26}_{-0.27}$ \\[4pt]
& (2.92$^{+0.07}_{-0.29}$) & (2.83$^{+0.20}_{-0.20}$) & (2.87$^{+0.20}_{-0.25}$) & (2.84$^{+0.19}_{-0.24}$) & (2.85$^{+0.21}_{-0.19}$) & (2.87$^{+0.27}_{-0.28}$) & (2.85$^{+0.20}_{-0.23}$) & (2.86$^{+0.20}_{-0.24}$) \\
\\

\multirow{2}*[-0.5ex]{$[$MgFe$]$\arcmin\tablenotemark{e} (\AA)}
& 3.12$^{+0.36}_{-0.38}$ & 3.34$^{+0.19}_{-0.22}$ & 3.25$^{+0.29}_{-0.25}$ & 3.35$^{+0.18}_{-0.41}$ & 3.20$^{+0.32}_{-0.25}$ & 3.45$^{+0.02}_{-0.47}$ & 3.28$^{+0.25}_{-0.32}$ & 3.24$^{+0.26}_{-0.34}$ \\[4pt]
& (3.49$^{+0.15}_{-0.28}$) & (3.43$^{+0.21}_{-0.28}$) & (3.47$^{+0.23}_{-0.26}$) & (3.47$^{+0.21}_{-0.27}$) & (3.50$^{+0.24}_{-0.32}$) & (3.46$^{+0.32}_{-0.32}$) & (3.47$^{+0.22}_{-0.28}$) & (3.35$^{+0.23}_{-0.31}$) \\
\\

\multirow{2}*[-0.5ex]{Mgb\tablenotemark{f} (\AA)}
& 3.85$^{+0.64}_{-0.29}$ & 4.03$^{+0.37}_{-0.38}$ & 3.91$^{+0.47}_{-0.54}$ & 4.09$^{+0.33}_{-0.73}$ & 3.81$^{+0.57}_{-0.53}$ & 3.86$^{+0.20}_{-0.45}$ & 4.02$^{+0.40}_{-0.66}$ & 3.83$^{+0.44}_{-0.59}$ \\[4pt]
& (4.33$^{+0.22}_{-0.40}$) & (4.26$^{+0.41}_{-0.51}$) & (4.32$^{+0.44}_{-0.44}$) & (4.33$^{+0.36}_{-0.40}$) & (4.40$^{+0.25}_{-0.58}$) & (4.29$^{+0.17}_{-0.54}$) & (4.32$^{+0.36}_{-0.49}$) & (4.02$^{+0.39}_{-0.56}$) \\
\\

\multirow{2}*[-0.5ex]{log($M_{BH}$/$M_{\odot}$)\tablenotemark{g}}
& 7.29$^{+0.76}_{-0.29}$ & 7.81$^{+0.23}_{-0.77}$ & 7.58$^{+0.55}_{-0.35}$ & 7.88$^{+0.29}_{-0.55}$ & 7.79$^{+0.49}_{-0.49}$ & 7.87$^{+0.23}_{-0.18}$ & 7.71$^{+0.45}_{-0.54}$ & 7.38$^{+0.44}_{-0.68}$ \\[4pt]
& (8.10$^{+0.30}_{-0.50}$) & (8.00$^{+0.36}_{-0.60}$) & (8.13$^{+0.30}_{-0.52}$) & (8.08$^{+0.32}_{-0.39}$) & (8.07$^{+0.31}_{-0.33}$) & (7.96$^{+0.42}_{-0.34}$) & (8.07$^{+0.33}_{-0.49}$) & (7.51$^{+0.51}_{-0.57}$) \\
\\

\multirow{2}*[-0.5ex]{H$\alpha$\tablenotemark{h} (\AA)}
& 0.52$^{+0.36}_{-0.52}$ & 0.48$^{+0.88}_{-0.42}$ & 0.61$^{+0.94}_{-0.44}$ & 0.48$^{+0.77}_{-0.33}$ & 0.41$^{+0.60}_{-0.35}$ & 0.81$^{+0.63}_{-0.62}$ & 0.49$^{+0.84}_{-0.38}$ & 0.49$^{+1.03}_{-0.34}$ \\[4pt]
& (0.33$^{+1.36}_{-0.28}$) & (0.35$^{+0.80}_{-0.27}$) & (0.35$^{+0.90}_{-0.26}$) & (0.38$^{+0.75}_{-0.31}$) & (0.53$^{+0.94}_{-0.39}$) & (0.63$^{+0.75}_{-0.43}$) & (0.39$^{+0.92}_{-0.31}$) & (0.45$^{+0.95}_{-0.33}$)

\enddata
%\tablecomments{}
\tablenotetext{a}{Effective velocity dispersion, velocity dispersion provided by the OSSY catalog with aperture corrections using the formulae in \citet{gra05} and \citet{cap06}.}
\tablenotetext{b}{Equivalent width (EW) of H$\beta$ absorption line provided by the OSSY catalog.}
\tablenotetext{c}{EW of Fe5270 absorption line.}
\tablenotetext{d}{EW of mean Fe absorption line, $<$Fe$>$ = (Fe5270+Fe5335)/2 \citep{gor90}}
\tablenotetext{e}{EW of $[$MgFe$]$\arcmin, $[$MgFe$]$\arcmin = $\sqrt{\rm Mgb(0.72 \times Fe5270 + 0.28 \times Fe5335)}$ \citep{tho03}}
\tablenotetext{f}{EW of Mgb absorption line.}
\tablenotetext{g}{Logarithmic scale of the central black mass derived using the formula in \citet{gul09}.}
\tablenotetext{h}{EW of H$\alpha$ emission line for non-AGN galaxies.}

\label{tab:catalogthree}
\end{deluxetable*}

%----------Summary Table 4
\setlength{\tabcolsep}{0.1cm}
\begin{deluxetable*}{lccccccccc}
\tablecaption{Basic properties of the Late-type and Irregular galaxies}
\tablewidth{0pt}
\tablehead{
\colhead{} &
\colhead{Sa} &
\colhead{Sb} &
\colhead{Sc} &
\colhead{Sd} &
\colhead{SBa} &
\colhead{SBb} &
\colhead{SBc} &
\colhead{SBd} &
\colhead{Irr}}

\startdata
\multirow{2}*[-0.5ex]{$\sigma_{eff}$ (\kms)}
& 114.0$^{+29.3}_{-27.4}$ & 61.9$^{+32.4}_{-20.1}$ & 48.8$^{+18.9}_{-11.5}$ & 40.4$^{+13.9}_{-10.7}$ & 101.8$^{+27.4}_{-31.0}$ & 75.5$^{+31.5}_{-19.0}$ & 53.2$^{+31.3}_{-14.3}$ &  0.0$^{+ 0.0}_{- 0.0}$ & 51.1$^{+10.9}_{- 9.9}$ \\[4pt]
& (122.3$^{+44.5}_{-33.5}$) & (69.8$^{+33.5}_{-25.8}$) & (52.1$^{+20.6}_{-15.1}$) & (57.2$^{+ 9.6}_{- 9.6}$) & (105.4$^{+37.1}_{-30.3}$) & (92.1$^{+23.5}_{-31.8}$) & (53.6$^{+46.1}_{-10.1}$) & (38.1$^{+ 0.0}_{- 0.0}$) & (42.2$^{+ 8.1}_{- 2.3}$) \\
\\

\multirow{2}*[-0.5ex]{H$\beta$ (\AA)}
& 2.25$^{+0.68}_{-0.40}$ & 2.90$^{+0.70}_{-0.69}$ & 2.93$^{+0.71}_{-0.61}$ & 3.98$^{+0.34}_{-0.70}$ & 2.37$^{+0.72}_{-0.52}$ & 2.64$^{+0.51}_{-0.73}$ & 3.13$^{+0.91}_{-0.88}$ & 4.41$^{+0.00}_{-0.00}$ & 4.01$^{+0.42}_{-0.31}$ \\[4pt]
& (2.11$^{+0.68}_{-0.40}$) & (2.84$^{+0.67}_{-0.67}$) & (3.14$^{+0.61}_{-0.68}$) & (4.14$^{+0.33}_{-1.14}$) & (2.16$^{+0.90}_{-0.42}$) & (2.43$^{+0.89}_{-0.54}$) & (3.14$^{+0.90}_{-0.88}$) & (4.33$^{+0.32}_{-0.36}$) & (4.18$^{+0.56}_{-0.22}$) \\
\\

\multirow{2}*[-0.5ex]{Fe5270 (\AA)}
& 2.56$^{+0.33}_{-0.49}$ & 2.14$^{+0.44}_{-0.66}$ & 1.95$^{+0.59}_{-0.65}$ & 1.06$^{+0.80}_{-0.22}$ & 2.48$^{+0.42}_{-0.70}$ & 2.31$^{+0.49}_{-0.67}$ & 1.79$^{+0.79}_{-0.60}$ & 1.12$^{+0.00}_{-0.00}$ & 1.24$^{+0.63}_{-0.44}$ \\[4pt]
& (2.64$^{+0.32}_{-0.46}$) & (2.12$^{+0.49}_{-0.63}$) & (1.89$^{+0.62}_{-0.59}$) & (1.07$^{+0.18}_{-0.82}$) & (2.61$^{+0.24}_{-0.60}$) & (2.46$^{+0.46}_{-0.89}$) & (1.77$^{+0.73}_{-0.80}$) & (1.29$^{+0.01}_{-0.32}$) & (1.06$^{+0.28}_{-0.55}$) \\
\\

\multirow{2}*[-0.5ex]{$<$Fe$>$ (\AA)}
& 2.56$^{+0.34}_{-0.44}$ & 2.18$^{+0.43}_{-0.64}$ & 2.02$^{+0.57}_{-0.52}$ & 1.01$^{+0.80}_{-0.17}$ & 2.49$^{+0.40}_{-0.65}$ & 2.42$^{+0.41}_{-0.61}$ & 1.98$^{+0.66}_{-0.44}$ & 0.50$^{+0.00}_{-0.00}$ & 1.24$^{+0.62}_{-0.56}$ \\[4pt]
& (2.67$^{+0.32}_{-0.48}$) & (2.16$^{+0.46}_{-0.59}$) & (1.94$^{+0.61}_{-0.50}$) & (0.84$^{+0.21}_{-0.50}$) & (2.62$^{+0.24}_{-0.59}$) & (2.46$^{+0.48}_{-0.85}$) & (2.10$^{+0.64}_{-0.42}$) & (0.83$^{+0.00}_{-0.00}$) & (0.84$^{+0.51}_{-0.34}$) \\
\\

\multirow{2}*[-0.5ex]{$[$MgFe$]$\arcmin (\AA)}
& 2.82$^{+0.40}_{-0.71}$ & 2.18$^{+0.52}_{-0.59}$ & 2.03$^{+0.48}_{-0.46}$ & 1.20$^{+0.24}_{-0.33}$ & 2.74$^{+0.47}_{-0.89}$ & 2.52$^{+0.56}_{-0.60}$ & 2.03$^{+0.75}_{-0.61}$ & 0.50$^{+0.00}_{-0.00}$ & 1.22$^{+0.25}_{-0.45}$ \\[4pt]
& (3.03$^{+0.43}_{-0.56}$) & (2.20$^{+0.53}_{-0.59}$) & (1.89$^{+0.63}_{-0.50}$) & (0.61$^{+0.13}_{-0.19}$) & (2.97$^{+0.39}_{-0.83}$) & (2.63$^{+0.60}_{-1.12}$) & (2.07$^{+0.59}_{-0.45}$) & (0.88$^{+0.00}_{-0.00}$) & (0.92$^{+0.24}_{-0.25}$) \\
\\

\multirow{2}*[-0.5ex]{Mgb (\AA)}
& 3.17$^{+0.65}_{-0.92}$ & 2.21$^{+0.73}_{-0.68}$ & 2.03$^{+0.64}_{-0.69}$ & 1.19$^{+0.25}_{-0.32}$ & 3.07$^{+0.72}_{-1.11}$ & 2.59$^{+0.83}_{-0.84}$ & 1.66$^{+1.03}_{-0.69}$ & 0.75$^{+0.00}_{-0.00}$ & 1.16$^{+0.13}_{-0.34}$ \\[4pt]
& (3.42$^{+0.76}_{-0.75}$) & (2.29$^{+0.82}_{-0.79}$) & (1.83$^{+0.71}_{-0.65}$) & (0.70$^{+0.17}_{-0.65}$) & (3.27$^{+0.79}_{-1.11}$) & (2.74$^{+1.00}_{-1.32}$) & (1.55$^{+0.87}_{-0.55}$) & (1.03$^{+0.16}_{-0.08}$) & (0.91$^{+0.21}_{-0.35}$) \\
\\

\multirow{2}*[-0.5ex]{log($M_{BH}$/$M_{\odot}$)}
& 7.10$^{+0.42}_{-0.50}$ & 5.86$^{+0.82}_{-0.90}$ & 5.25$^{+0.72}_{-1.29}$ & 4.57$^{+1.02}_{-1.11}$ & 6.87$^{+0.43}_{-0.67}$ & 6.25$^{+0.70}_{-1.02}$ & 5.06$^{+1.30}_{-1.38}$ & 4.64$^{+0.00}_{-0.00}$ & 5.22$^{+0.54}_{-1.25}$ \\[4pt]
& (7.20$^{+0.59}_{-0.64}$) & (6.02$^{+0.84}_{-0.95}$) & (5.30$^{+0.82}_{-1.41}$) & (4.68$^{+1.01}_{-1.45}$) & (6.94$^{+0.55}_{-0.62}$) & (6.59$^{+0.51}_{-1.02}$) & (4.49$^{+1.22}_{-0.72}$) & (4.37$^{+0.51}_{-0.37}$) & (4.36$^{+0.90}_{-0.68}$) \\

\\
\multirow{2}*[-0.5ex]{H$\alpha$ (\AA)}
& 4.44$^{+18.62}_{- 3.32}$ & 12.36$^{+13.27}_{- 7.76}$ & 12.50$^{+11.42}_{- 6.93}$ & 27.72$^{+20.23}_{-10.22}$ & 5.87$^{+23.52}_{- 4.79}$ & 13.57$^{+16.00}_{-11.17}$ & 19.95$^{+20.52}_{-12.79}$ & 55.04$^{+ 0.00}_{- 0.00}$ & 59.09$^{+39.09}_{-26.71}$ \\[4pt]
& (1.66$^{+10.65}_{- 1.01}$) & (12.29$^{+16.04}_{- 8.61}$) & (16.53$^{+17.70}_{- 9.31}$) & (54.98$^{+18.12}_{-21.88}$) & (2.11$^{+24.49}_{- 1.37}$) & (11.27$^{+27.92}_{-10.00}$) & (23.21$^{+19.05}_{- 8.78}$) & (38.15$^{+10.54}_{- 5.47}$) & (47.28$^{+32.96}_{-20.63}$)

\enddata

\label{tab:catalogfour}
\end{deluxetable*}

% --------------------------------------------------------------------------------

\begin{deluxetable*}{cccccc}
\tablecolumns{12}
%\tabletypesize{\scriptsize}
\tablecaption{Morphology Catalog Including Photometric and Spectroscopic  Properties}
\tablewidth{0pt}
\tablehead{
\colhead{SDSS ObjID} &
\colhead{Morphology\tablenotemark{a}} &
\colhead{R.A.\tablenotemark{b}} (deg) &
\colhead{Dec.\tablenotemark{b} (deg)} &
\colhead{Redshift} &
\colhead{$(g-r)_{0}$} 
}
\startdata
587739828207353983  &    13   &    228.61371  &     20.78934  &   0.0397760   &   0.780  \\
588017703493500993  &    21   &    240.40990  &     7.10945  &    0.0400339  &    0.723  \\
587726100949434515  &    16   &    220.11459  &     3.68553  &    0.0279422   &   0.713      \\ 
587739407337455890  &    23   &    232.81613  &     25.31393  &   0.0341272   &   0.490     \\
587739829271855250  &    23   &    206.47068  &     26.77517  &   0.0298094   &   0.521   \\ [2pt]
\hline

$M_{\rm r}$ & $\rm IsoA_r$ (arcsec) & $\rm IsoB_r$ (arcsec) & $C_{r}$ & \fracDeV  & log($M_{\ast}$/$M_{\odot}$) \\ [2pt]
\hline
$-22.182$   &    63.033   &   46.404   &   3.321   &   1.000   &   11.290   \\
$-21.198$  &    38.271   &   26.423   &   3.013   &   0.983   &   10.834   \\
$-20.453$   &    66.449   &   24.693   &   2.598   &   0.610   &   10.525   \\
$-20.223$  &    44.662   &   34.365   &   2.134   &   0.105   &   10.189   \\
$-21.061$   &    66.997   &   32.829   &   2.082   &   0.000   &   10.558    \\ [2pt]
\hline
$\sigma_{eff}$\tablenotemark{c} (\kms) & $error(\sigma_{eff}$)\tablenotemark{c} (\kms) & H$\beta$\tablenotemark{d} (\AA) & Fe5270\tablenotemark{d} (\AA) & $<$Fe$>$\tablenotemark{d}  (\AA)& Mgb\tablenotemark{d} (\AA) \\ [2pt]
\hline
293.879   &   3.362   &   1.658   &   3.045   &   3.005   &    5.072   \\ 
109.202   &   2.331   &   2.233   &   2.955   &   2.656   &    3.312   \\
76.974    &   3.516   &   2.265   &   2.880   &   2.688   &    2.608   \\
33.622    &   9.431   &   2.217   &   1.427   &   1.243   &    2.696   \\
59.220    &   4.451   &   2.503   &   1.663   &   1.313   &    2.471   \\ [2pt]
\hline

$[$MgFe$]$\arcmin\tablenotemark{d} (\AA) & age\tablenotemark{e} (Gyr) & [Z/H]\tablenotemark{e} & [$\alpha$/Fe]\tablenotemark{e} & BPT class\tablenotemark{f} & log($M_{BH}$/$M_{\odot}$)\tablenotemark{g} \\ [2pt]
\hline
3.915   &   10.789   &   0.321   &   0.363   &   4   &   8.829   \\
3.038   &   3.111    &   0.262   &   0.328   &   5   &   7.006   \\
2.689   &   3.062    &   0.097   &   0.120   &   5   &   6.362   \\
1.889   &   8.056    &   $-0.610$  &   $-0.504$  &   5   &   4.836   \\
1.904   &   4.704    &   $-0.482$  &   $-0.352$  &   1   &   5.879   \\[2pt]
\hline

H$\alpha$\tablenotemark{h} $($\AA$)$ & N$_{companion}$\tablenotemark{i} \\ [2pt]
\hline
$-9999$   &   2   \\
$-9999$   &   0   \\
0.299   &   8   \\
4.618   &   3   \\
9.615   &   0
\enddata

\label{tab:tablesix}
\tablecomments{5 example galaxies among 6,069 objects in the catalog of isolated and comparison sample galaxies.}
\tablenotetext{a}{E0--E5: 10--15, S0: 16, Sa--Sd: 21--24, SBa--SBd: 31--34, Irr: 40, unknown: 50.}
\tablenotetext{b}{J2000}
\tablenotetext{c}{`$-9999$' is assigned when a galaxy does not satisfy 10 $<$ $\sigma_{eff}$ $<$ 400 \kms.}
\tablenotetext{d}{`$-9999$' is assigned when a galaxy does not satisfy $S/N > 3$ and N$\sigma<2$.}
\tablenotetext{e}{Properties derived from the stellar population model, valid for statistical studies like Figure~\ref{fig:age}--\ref{fig:a_fe}, however may differ more significantly for galaxies on a one-to-one basis. `$-9999$' is assigned when a galaxy does not satisfy $S/N > 3$ and N$\sigma<2$.}
\tablenotetext{f}{Weak-emission: 0, Star-forming: 1, Composite: 2, Seyfert: 3, LINER: 4, Unclear: 5.}
\tablenotetext{g}{`$-9999$' is assigned when $\sigma_{eff}$ is `$-9999$'.}
\tablenotetext{h}{`$-9999$' is assigned for AGN host galaxies or negative values.}
\tablenotetext{i}{The number of companion galaxies within 1 Mpc in three dimensional space. 0 indicates isolated galaxies and $\geq$2 means comparison galaxies.}
\end{deluxetable*}

\end{document}